\documentclass[a4paper,twocolumn,english,prb,aps,showpacs,floatfix,groupedaddress,superscriptaddress]{revtex4-1}
\usepackage{graphicx}
\usepackage{epsfig}
\usepackage[english]{babel}
\usepackage{amsmath}
\usepackage{amssymb}
\usepackage{amsfonts}
\usepackage{longtable}
\usepackage{dcolumn}
\usepackage{bm}
\usepackage{bbm}
\usepackage{nicefrac}
\usepackage{color,array}
\usepackage{colortbl}
\usepackage{dsfont}
\usepackage{mathtools}
\usepackage{subfigure}  
\usepackage{soul}

\usepackage{xcolor,cancel}

\usepackage[breaklinks,colorlinks=true,citecolor=blue,linkcolor=blue]{hyperref}
\usepackage{breakurl}

\begin{document}

\newcommand{\be}{\begin{equation}}
\newcommand{\ee}{\end{equation}}
\newcommand{\bearr}{\begin{eqnarray}}
\newcommand{\eearr}{\end{eqnarray}}
\newcommand{\bseq}{\begin{subequations}}
\newcommand{\eseq}{\end{subequations}}
\newcommand{\nn}{\nonumber}
\newcommand{\dagg}{{\dagger}}
\newcommand{\vpdag}{{\vphantom{\dagger}}}
\newcommand{\bpm}{\begin{pmatrix}} 
\newcommand{\epm}{\end{pmatrix}} 
\newcommand{\bs}{\boldsymbol}

\title{Artificial $SU(3)$ Spin-Orbit Coupling and Exotic Mott Insulators}

\author{Mohsen Hafez-Torbati}
\email{torbati@itp.uni-frankfurt.de}
\affiliation{Institut f{\"u}r Theoretische Physik, Goethe-Universit{\"a}t,
60438 Frankfurt/Main, Germany.}

\author{Walter Hofstetter}
\email{hofstett@physik.uni-frankfurt.de}
\affiliation{Institut f{\"u}r Theoretische Physik, Goethe-Universit{\"a}t,
60438 Frankfurt/Main, Germany.}

\date{\today}

\begin{abstract}
Motivated by recent progress in the realization of artificial gauge fields and 
$SU(N)$ Mott insulators using alkaline-earth-like atoms in optical lattices, we 
develop an unbiased $SU(N)$ real-space dynamical mean-field theory (DMFT) approach to study the 
effect of spin-orbit coupling and onsite Hubbard interaction $U$ on $SU(3)$ fermionic 
systems. We investigate the behavior of the 
local magnetization, double occupancies, and the triple 
occupancy versus the Hubbard interaction across the metal to Mott insulator transition. We map 
out the magnetic phase diagram in the large-$U$ limit and show that the spin-orbit coupling 
can stabilize long-range orders such as ferromagnet, spiral, and stripes with different 
orientations in $SU(3)$ Mott insulators.
\end{abstract}


\maketitle

\section{Introduction}
Since the experimental observation of Bose-Einstein condensation \cite{Anderson1995}, 
ultracold atoms have attracted a lot of attention as a 
flexible playground to mimic various models of condensed matter physics \cite{Bloch2005} 
and beyond \cite{Georgescu2014}. The Haldane 
model is a fundamental model in the field of topological insulators which describes the transition between 
topologically distinct phases in the absence of net magnetic flux through the unit cell \cite{Haldane1988}. 
Thanks to the lattice-shaking technique and Floquet theory, the Haldane model which was initially considered 
difficult to realize is implemented in optical lattices and its phase diagram has been mapped out \cite{Jotzu2014}. 
In the same spirit, although technically differently, the time-reversal-invariant Hofstadter Hamiltonian 
has been realized \cite{Aidelsburger2011,Aidelsburger2013,Miyake2013}, 
the Chern number of the lowest Hofstadter band has been determined \cite{Aidelsburger2014}, 
and the Berry curvature of the Bloch bands 
measured \cite{Flaschner2016}. Going beyond Abelian gauge fields \cite{Jaksch2003}, non-Abelian gauge fields 
for $SU(2)$ systems can also be engineered in optical lattices 
\cite{Osterloh2005,Satija2006,Goldman2010} leading, for example, to the prediction of 
the Hofstadter moth \cite{Osterloh2005}. 

The perfect decoupling of nuclear spin from electronic angular momentum in alkaline-earth atoms 
provides a unique possibility to study $SU(N)$ Mott insulators in optical lattices with $N$ as large as $10$ \cite{Wu2003,Honerkamp2004,Gorshkov2010a}. 
Depending on the value of $N$ and the lattice geometry, phases such as multi-flavor magnetism \cite{Toth2010,Sotnikov2014}, 
valence-bond solid states \cite{Hermele2009,Hermele2011,Zhou2016}, and quantum spin liquids \cite{Hermele2009,Hermele2011} are predicted to emerge. 
While the strong spin-orbit coupling in $SU(2)$ Mott insulators leads 
to phases such as collinear, spiral, and tetrahedral spin orders \cite{Cole2012,Radic2012,Cocks2012,Hickey2015},
in high-spin systems the phase diagram is expected to be richer 
not only in magnetic order but also with respect to topology. For instance, a translationally constant gauge field in 
an $SU(3)$ system can lead to non-trivial topological bands, in contrast to the $SU(2)$ case \cite{Barnett2012}. 
Proposals to realize effective $SU(3)$ spin-orbit coupling in optical lattices already exist \cite{Juzeliunas2010,Dalibard2011,Goldman2014}.

In this paper, we explore the $SU(3)$ Hubbard model on the triangular lattice at $1/3$-filling 
subject to homogeneous non-Abelian 
gauge fields, using the real-space dynamical mean-field theory (DMFT) approximation \cite{Song2008,Snoek2008}. 
The method is applied to various 2-component models with spin-orbit coupling \cite{Cocks2012,Irsigler2018,Zheng2018}. 
We have employed the exact diagonalization (ED) approach as 
the impurity solver. 
Although the $SU(3)$ Hubbard model without gauge fields has been investigated near $1/2$-filling with anisotropic interactions
\cite{Inaba2010,Miyatake2010,Inaba2012} and  
at $1/3$-filling in the large-$U$ limit \cite{Toth2010,Bauer2012,Sotnikov2014},
there are no results available across the metal to Mott insulator transition 
in the $SU(3)$-symmetric version.
First, we set the gauge field to zero and show that the $SU(3)$ Hubbard model on the triangular lattice at $1/3$-filling 
shows a transition from a metallic phase to Mott insulator with 3-sublattice magnetic order at the Hubbard interaction 
$U_c \simeq 10.7t$. We study the behavior of different local quantities such as the magnetization, double occupancies, 
and the triple occupancy versus $U$. Next, we analyze the effect of gauge fields 
on the emergence of exotic $SU(3)$ magnetism in the Mott regime. We find $SU(3)$ Mott insulators 
with long-range orders 
such as ferromagnetic, spiral, and stripes with different kinds of orientations.

\section{Technical Aspects}
\label{sec:dmft}
The real-space DMFT (RDMFT) method was initially introduced to investigate film geometries, where a large but finite number 
of layers are coupled \cite{Potthoff1999}. Since then, RDMFT has been applied to different problems \cite{Hofstetter2018} ranging  
from disordered systems \cite{Song2008,Zheng2018} to topological insulators \cite{Cocks2012,Irsigler2018} and exotic magnetism \cite{Orth2013,He2015}.

We adapt the RDMFT method in this section to address $SU(N)$ systems in the presence of flavor-mixing hopping terms. A hopping 
term which flips the spin can be induced as a result of spin-orbit coupling in solid state systems or by creating 
artificial gauge-fields for ultracold atoms in optical lattices \cite{Goldman2014}. Due to the recent progress in 
realization of $SU(N)$ systems and artificial gauge fields in optical lattices, such a methodological development 
seems in high demand.

Here we consider the version of RDMFT in which the self-energy $\bs{\Sigma}(i\omega_n)$ is approximated to be spatially local but it can be 
position dependent, 
\be
\left[\bs{\Sigma}(i\omega_n) \right]^\vpdag_{\bs{r}\alpha,\bs{r}'\alpha'}=\delta_{\bs{r}\bs{r}'}^\vpdag \left[\bs{\Sigma}(i\omega_n) \right]^\vpdag_{\bs{r}\alpha,\bs{r}\alpha'},
\ee
where $\bs{r}$ specifies a lattice position, $\alpha$ and $\alpha'$ are internal degrees of freedom, 
$\delta_{\bs{r}\bs{r}'}$ is the Kronecker delta function, and $\omega_n$ stands for a Matsubara frequency. 
The notation $\left[ M \right]_{m,n}$ is used to refer to the elements of the matrix $M$.
We notice that the self-energy matrix $\bs{\Sigma}(i\omega_n)$ is block-diagonal with the size of each block being $N\times N$ for 
an $SU(N)$ system.

The self-consistency cycle starts with an initial guess for the self-energy, from which the lattice Green function $\bs{G}(i\omega_n)$ can be computed 
using the lattice Dyson equation:
\be 
\bs{G}(i\omega_n) = \left[ i\omega_n \mathds{\bs{1}} - \bs{H}_0 -\bs{\Sigma}(i\omega_n)\right]^{-1},
\label{eq:latDyson}
\ee
where $\bs{H}_0$ is the matrix representation of the model Hamiltonian $H$ in the 1-particle 
subspace $\left\{ | \bs{r} \alpha \rangle \right\}$:
\be
\left[ \bs{H}_0\right]^\vpdag_{\bs{r}\alpha,\bs{r}'\alpha'}:=\langle \bs{r} \alpha | H | \bs{r}' \alpha' \rangle \quad.
\ee
Using Eq. \eqref{eq:latDyson} one finds the local Green function $\bs{\mathcal{G}^\vpdag_{\bs{r}}}(i\omega_n)$ 
which is an $N\times N$ matrix given by 
\be 
\left[\bs{\mathcal{G}_{\bs{r}}}(i\omega_n) \right]^\vpdag_{\alpha,\alpha'}:=\left[\bs{G}(i\omega_n) \right]^\vpdag_{\bs{r}\alpha,\bs{r}\alpha'},
\ee
and subsequently the inverse dynamical Weiss field $\bs{\mathcal{G}}^{(0)}_{\bs{r}}(i\omega_n)^{-1}$ is calculated from 
the local Dyson equation
\be
\bs{\mathcal{G}}^{(0)}_{\bs{r}}(i\omega_n)^{-1}= \bs{\mathcal{G}_{\bs{r}}^\vpdag}(i\omega_n)^{-1}+\bs{\Sigma_{\bs{r}}^\vpdag}(i\omega_n)
\label{eq:weiss}
\ee
with $\left[\bs{\Sigma}_{\bs{r}}(i\omega_n) \right]^\vpdag_{\alpha,\alpha'}:=\left[\bs{\Sigma}(i\omega_n) \right]^\vpdag_{\bs{r}\alpha,\bs{r}\alpha'}$.
We consider the $SU(N)$ Anderson impurity model (AIM)
\bearr
H_{\bs{r}}^{\rm AIM}=&-&  \Psi^\dagg_{\bs{r}} \bs{\mu}^\vpdag_{\bs{r}} \Psi^\vpdag_{\!\bs{r}}
+\sum_{\alpha<\alpha'} U_{\alpha\alpha'}^{\bs{r}} n^\vpdag_{\bs{r}\alpha} n^\vpdag_{\bs{r}\alpha'} \nn \\
&+&\sum_{l=1}^{l_{\rm max}} \varepsilon_l^{\bs{r}} \Phi^\dagg_{l} \Phi^\vpdag_{l}
+\sum_{l=1}^{l_{\rm max}} \left( \Phi^\dagg_{l} \bs{V}_l^{\bs{r}} \Psi^\vpdag_{\bs{r}} + {\rm H.c.} \right)
\label{eq:aim}
\eearr
to describe the local physics at the lattice position $\bs{r}$. The chemical potential matrix $\bs{\mu}^\vpdag_{\bs{r}}$ 
is defined as $\left[\bs{\mu}_{\bs{r}}\right]_{\alpha,\alpha'}:=-\langle \bs{r} \alpha | H | \bs{r} \alpha \rangle \delta^\vpdag_{\alpha\alpha'}$, and 
$U_{\alpha\alpha'}^{\bs{r}}$ is the Hubbard interaction between flavors $\alpha$ and $\alpha'$ at the lattice site $\bs{r}$.
The $SU(N)$ field operators $\Psi^\vpdag_{\!\bs{r}}$ and $\Phi^\vpdag_{l}$ act at the lattice 
position $\bs{r}$ and at the bath orbital $l$, respectively. They are column vectors with the elements $\left[\Psi_{\!\bs{r}}\right]_\alpha=c^\vpdag_{\bs{r}\alpha}$, 
$\left[\Phi_{l}\right]_\alpha=a^\vpdag_{l\alpha}$, where $c^\vpdag_{\bs{r}\alpha}$ and $a^\vpdag_{l\alpha}$ are the normal fermionic annihilation operators at the 
impurity site $\bs{r}$ and at the bath orbital $l$ with the flavor $\alpha$. We have also defined 
$n_{\bs{r}\alpha}^\vpdag=c^\dag_{\bs{r}\alpha}c^\vpdag_{\bs{r}\alpha}$. 
The real parameters $\varepsilon_l^{\bs{r}}$ describe 
the bath onsite energies and the $N\times N$ matrices $\bs{V}_l^{\bs{r}}$ with complex elements describe the hopping from the impurity 
to the bath. They are determined by fitting the dynamical Weiss field \eqref{eq:weiss} to the finite-orbital function
\be 
\tilde{\bs{\mathcal{G}}}^{(0)}_{\bs{r}}(i\omega_n)^{-1}= i\omega_n \mathds{1} + \bs{\mu}_{\bs{r}}
-\sum_{l=1}^{l_{\rm max}} \frac{{\bs{V}_l^{\bs{r}}}^{\dag} \bs{V}_l^{\bs{r}} }{i\omega_n-\varepsilon_l^{\bs{r}}}
\ee
via a least-square minimization process. The AIM is diagonalized exactly 
and the finite-orbital interacting Green function $\tilde{\bs{\mathcal{G}}}_{\bs{r}}(i\omega_n)$ at the 
impurity site is obtained using the Lehmann representation. The new self-energy is calculated via
\be
\bs{\Sigma_{\bs{r}}^\vpdag}(i\omega_n)= \tilde{\bs{\mathcal{G}}}^{(0)}_{\bs{r}}(i\omega_n)^{-1}-\tilde{\bs{\mathcal{G}}}_{\bs{r}}^\vpdag(i\omega_n)^{-1},
\ee
and is used for the next iteration. 

The matrix inversion \eqref{eq:latDyson} and the AIM diagonalization are 
the two main time-consuming parts of the RDMFT and in both cases exploiting the symmetry  
reduces the runtime significantly. Using the translational symmetry of the 
phase under study we consider the nonequivalent lattice sites closest to the lattice 
center as ``representative sites'', for which the corresponding columns of the lattice Green function \eqref{eq:latDyson} 
are found and the AIM \eqref{eq:aim} is set up. The representative sites are chosen to be close to the 
lattice center in order to minimize the edge effects on the bulk properties in the case of open boundary conditions. 
We need to keep track of the self-energy $\bs{\Sigma}^\vpdag_{\bs{r}}(i\omega_n)$, the inverse dynamical 
Weiss field $\bs{\mathcal{G}}^{(0)}_{\bs{r}}(i\omega_n)^{-1}$, and the local interacting Green function 
$\bs{\mathcal{G}}^\vpdag_{\bs{r}}(i\omega_n)$ solely at the representative sites. 
Only for the construction of the inverse lattice Green function in Eq. \eqref{eq:latDyson} 
one needs to temporarily generate the self-energy over the full lattice. This full exploitation of the translational symmetry allows 
us to address translationally ordered and disordered systems efficiently on an equal footing 
using RDMFT.

We notice that as far as the hopping terms in the Hamiltonian $H$ are of short range the matrix 
$\bs{H}_0$ can be treated as an sparse matrix, and if in addition the boundary conditions are open 
(one or all of them) it can be realized as a block-tridiagonal matrix, which allow for a fast inversion.

In contrast to the normal DMFT method the above described approach is not biased towards 
any specific solution in the case of 
spontaneous breaking of $SU(N)$ symmetry, and, in principle, one can produce all the degenerate 
states. However, it is also possible to concentrate on a solution with a diagonal dynamical Weiss field 
and consider the hopping matrices $\bs{V}_l^{\bs{r}}$ to be diagonal in order to exploit the conservation 
of the total charge for {\it each} flavor in the diagonalization of the AIM \eqref{eq:aim}. 
An example would be the N\'eel AF order with spins pointing in the $\hat{S}_z$-direction 
in the $SU(2)$ Hubbard model on the square lattice. However, one notices that there is not such a solution 
in the $SU(2)$ Hubbard model on the triangular lattice where spins form a $120^\circ$ spiral order.

Now we discuss the calculation of the total energy for fermionic $SU(N)$ systems using the RDMFT method. 
While the contributions of the local terms to the energy can be easily found from the local impurity problem, computing 
the non-local contributions coming from the hopping term is a bit challenging. But it can also 
be done in an straightforward manner. We consider a general hopping term given by
\be 
H_t^\vpdag = \sum_{\bs{r} \bs{r}'} \Psi^\dagg_{\bs{r}} \bs{T}^\vpdag_{\bs{r}\bs{r}'} \Psi^\vpdag_{\bs{r}'}
\ee
where the $N\times N$ hopping matrix $\bs{T}^\vpdag_{\bs{r}\bs{r}'}$ satisfies 
$\bs{T}^\vpdag_{\bs{r}\bs{r}'}=\bs{T}^\dagg_{\bs{r}'\bs{r}}$ and we have supposed 
$\bs{T}^\vpdag_{\bs{r}\bs{r}'}=\bs{0}$ for $\bs{r}=\bs{r}'$. Using the imaginary time Green 
function and its Fourier transform one obtains 
\be 
\langle H_t \rangle \!=\! \lim_{\epsilon \to 0^+} \frac{1}{\beta} 
\sum_{n} \sum_{\bs{r} \bs{r}'} \sum_{\alpha \alpha'}  
\left[ \bs{T}_{\bs{r}\bs{r}'} \right]_{\alpha, \alpha'} e^{+i\omega_n \epsilon} 
\left[ \bs{G}(i\omega_n) \right]_{\bs{r}\alpha,\bs{r'}\alpha'}
\label{eq:kinetic_tmp}
\ee
where $\beta$ is the inverse temperature. Using the lattice Dyson equation \eqref{eq:latDyson} 
the hopping matrix can be expressed as
\bearr
\left[ \bs{T}_{\bs{r}\bs{r}'} \right]_{\alpha, \alpha'}=
\left[ \bs{\Delta_{\bs{r}}}(i\omega_n) \right]_{\alpha, \alpha'} \delta_{\bs{r} \bs{r'}}
&+&\left[ \bs{\mathcal{G}}_{\bs{r}}(i\omega_n)^{-1} \right]_{\alpha, \alpha'} \delta_{\bs{r} \bs{r'}} \nn \\
&-&\left[ \bs{G}(i\omega_n)^{-1} \right]_{\bs{r}\alpha, \bs{r'}\alpha'}
\label{eq:hoppingM}
\eearr
where we have used the local Dyson equation \eqref{eq:weiss} to substitute the self-energy and we have defined 
the hybridization function 
$\bs{\Delta}_{\bs{r}}(i\omega_n)=i\omega_n \mathds{1}+\bs{\mu}_{\bs{r}}-\bs{\mathcal{G}}^{(0)}_{\bs{r}}(i\omega_n)^{-1}$.
Substituting the hopping matrix from Eq. \eqref{eq:hoppingM} into Eq. \eqref{eq:kinetic_tmp} we get
\be 
\langle H_t \rangle \!=\! \lim_{\epsilon \to 0^+} \frac{1}{\beta} 
\sum_{n} \sum_{\bs{r} }   
e^{+i\omega_n \epsilon} 
{\rm Tr}\left[ \bs{\Delta}_{\bs{r}}(i\omega_n) \bs{\mathcal{G}}_{\bs{r}}(i\omega_n)  \right] ,
\label{eq:kinetic}
\ee
which expresses the kinetic energy in terms of only local functions. The value of $\epsilon$ 
in Eq. \eqref{eq:kinetic} can safely be set to zero as the summand falls off as $1/\omega_n^2$ for large $\omega_n$.
In practical calculations one requires to introduce a cutoff for Matsubara frequencies. The total energy 
of the system reads
\be
E= \langle H_t \rangle +\sum_{\bs{r}} {\rm Tr}\left[ \bs{\mu}_{\bs{r}} \bs{\rho}_{\bs{r}}^{T} \right]
+\frac{1}{2}\sum_{\bs{r}} {\rm Tr}\left[ \bs{U}_{\bs{r}} \bs{d}_{\bs{r}} \right]
\ee
where $\left[\bs{\rho}_{\bs{r}}\right]_{\alpha,\alpha'}:=\langle c^\dagg_{\bs{r}\alpha} c^\vpdag_{\bs{r}\alpha'} \rangle$, 
$\left[\bs{d}_{\bs{r}}\right]_{\alpha,\alpha'}:=\langle n^\vpdag_{\bs{r}\alpha} n^\vpdag_{\bs{r}\alpha'} \rangle$, and 
$\left[\bs{U}_{\bs{r}}\right]_{\alpha,\alpha'}:=U_{\alpha\alpha'}^{\bs{r}}=U_{\alpha'\alpha}^{\bs{r}}$ with the 
assumption $U_{\alpha\alpha}^{\bs{r}}=0$.
In the following we apply the above formalism to the $SU(3)$ Hubbard model on the triangular lattice with and without 
spin-orbit coupling and further applications of the method will be discussed in future publications.

\section{Hamiltonian}
\begin{figure}[t]
    \centering
     \includegraphics[width=0.82\linewidth,angle=0]{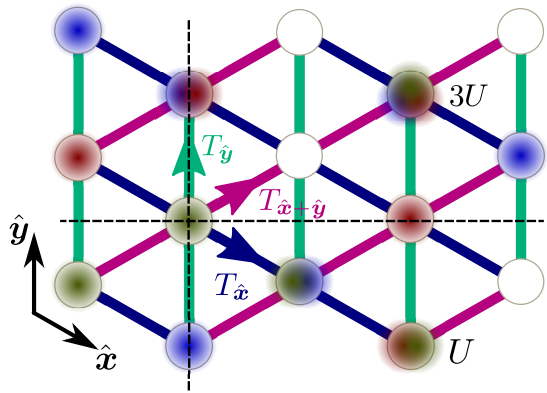}
     \caption{(color online). Schematic representation of the $SU(3)$ Hamiltonian \eqref{eq:hamiltonian} on 
     the triangular lattice. The hopping matrices in the directions $\hat{\bs x}$, $\hat{\bs x}+\hat{\bs y}$, and $\hat{\bs y}$ 
     are denoted by blue, red, and green bonds. The interaction between different flavors occupying the same lattice 
     site is described by the isotropic Hubbard interaction $U$. 
     The dashed lines 
     are guide-to-eye for a better realization of reflection symmetries with respect to the horizontal and vertical axes.}
     \label{fig:hamiltonian}
\end{figure}

We consider the Hamiltonian 
\bearr
\label{eq:hamiltonian}
H\!=&-&t\sum_{\bs r}\sum_{{\bs \delta}} 
\left(
\Psi^\dagg_{\bs{r}+\bs{\delta}} \bs{T}_{\bs \delta}^\vpdag \Psi^\vpdag_{\bs r} + {\rm H.c.}
\right)
+ U \sum_{\bs r}\!\sum_{\alpha<\alpha'}\! n^\vpdag_{\bs{r}\alpha} n^\vpdag_{\bs{r}\alpha'}  \nn \\
&-& \mu \sum_{\bs r} \Psi^\dagg_{\bs{r}} \Psi^\vpdag_{\bs{r}} \quad,
\eearr
where we have defined the $SU(3)$ creation field operator 
$\Psi^\dagg_{\bs r}=\left( c^\dagg_{{\bs r},+1}, c^\dagg_{{\bs r}, 0},c^\dagg_{{\bs r}, -1} \right)$. 
The operators $c^\dagg_{{\bs r} \alpha}$ and $c^\vpdag_{{\bs r} \alpha}$ represent the 
fermionic creation and annihilation operators at the position ${\bs r}$ with the flavor $\alpha=\pm 1,0$. 
We define also the occupation operator $n^\vpdag_{{\bs r} \alpha}=c^\dagg_{{\bs r} \alpha}c^\vpdag_{{\bs r} \alpha}$.
The position vector ${\bs r}$ runs over the triangular lattice and the nearest-neighbor (NN) vector 
can have the values ${\bs \delta}=\hat{\bs x},\hat{\bs y},\hat{\bs x}+\hat{\bs y}$ where $\hat{\bs x}$ and $\hat{\bs y}$ 
are the unit vectors according to the 
reference frame specified in Fig. \ref{fig:hamiltonian} with the lattice constant set to unity. The first term in Eq. \eqref{eq:hamiltonian} describes 
the NN hopping in the three different directions on the triangular lattice, see Fig. \ref{fig:hamiltonian}, 
the second term is the $SU(3)$-symmetric Hubbard interaction, and the last term enables us to reach the desired filling by adjusting 
the chemical potential $\mu$. The blue, red, and green bonds 
in Fig. \ref{fig:hamiltonian} correspond to the hopping in $\hat{\bs x}$, $\hat{\bs x}+\hat{\bs y}$, and $\hat{\bs y}$ directions, respectively. 
The diagonal elements of the hopping matrix $\bs{T}_{\bs \delta}$ depict flavor-conserving hoppings and its off-diagonal elements 
depict flavor-mixing hoppings. The hopping matrices in the three different directions are given by 
$\bs{T}_{\hat{\bs x}}=e^{+2\pi i \gamma (\bs{\lambda}_2-\bs{\lambda}_5+\bs{\lambda}_7)/\sqrt{3}}$, 
$\bs{T}_{\hat{\bs y}}=e^{+\pi i \kappa (\bs{\lambda}_3+\sqrt{3}\bs{\lambda}_8)}$, and 
$\bs{T}_{\hat{\bs x}+\hat{\bs y}}=e^{+\pi i \kappa} \bs{T}_{\hat{\bs x}}\bs{T}_{\hat{\bs y}}$ 
where $\bs{\lambda}_1 \cdots \bs{\lambda}_8$ are the eight Gell-Mann matrices and 
$\gamma$ and $\kappa$ are spin-orbit coupling constants. 
With this choice of hopping matrices, 
the first term in Eq. \eqref{eq:hamiltonian} at $\gamma=\kappa=1/3$ can be mapped by the gauge transformation (${\bs r}=x\hat{\bs x}+y\hat{\bs y}$)
\be 
\Psi^\dagg_{\bs r} \longrightarrow \Psi^\dagg_{\bs r} e^{+\pi i \kappa y (\bs{\lambda}_3+\sqrt{3}\bs{\lambda}_8)} \bs{\mathcal{U}}^x; ~ 
\bs{\mathcal{U}}= -i
\bpm
0 & 0 & 1 \\
1 & 0 & 0 \\
0 & 1 & 0
\epm,
\ee
to three identical copies of the Hatsugai-Harper-Hofstadter model described by 
$\bs{T}_{\hat{\bs x}}=\bs{\mathds{1}}$, $\bs{T}_{\hat{\bs x}+\hat{\bs y}}=e^{+2\pi i \phi (2x+1)}\bs{\mathds{1}}$, 
and $\bs{T}_{\hat{\bs y}}=e^{+4\pi i \phi x}\bs{\mathds{1}}$ with the flux per triangle $\phi=1/6$ \cite{Hatsugai1990}. Therefore, the 
Hamiltonian \eqref{eq:hamiltonian} at ${\gamma=\kappa=1/3}$ is equivalent to the $SU(3)$ Hatsugai-Harper-Hofstadter-Hubbard (HHHH) model, 
which at $1/3$ filling shows a $\mathcal{C}=-3$ Chern insulator at weak interaction and a Mott insulator in the large-$U$ limit. 

Despite the interesting topological properties of the Hamiltonian \eqref{eq:hamiltonian}, in this work we 
concentrate mainly on the exotic magnetic textures, which appear in the Mott regime. We would like to mention the recent analyses of 
the $SU(2)$ Haldane-Hubbard model, which identify not only interaction-induced Chern insulators at moderate interactions \cite{Vanhala2016}
but also interesting spin orders in the Mott insulator phase \cite{Hickey2015,Zheng2015}.
The topological properties of the Hamiltonian \eqref{eq:hamiltonian} in the non-interacting limit and 
at ${\gamma=\kappa=1/3}$ have been analyzed on the square lattice in Ref. \onlinecite{Barnett2012}, 
where the relation between $N$ decoupled copies of the Harper-Hofstadter model and an $SU(N)$  system 
with homogeneous non-Abelian gauge fields was first proven.

The Hamiltonian \eqref{eq:hamiltonian} for $\gamma=\kappa=0$ reduces to the $SU(3)$ symmetric Hubbard 
model which can be realized in optical lattices using fermionic $^6$Li atoms at large magnetic field, where 
nuclear spin and electronic angular momentum get decoupled \cite{Ottenstein2008,Huckans2009}.
The spin-orbit coupling at finite $\gamma$ and $\kappa$ can be realized by creating artificial gauge fields 
in optical lattices using, for example, the so-called tetrapod setup scheme \cite{Juzeliunas2010,Dalibard2011,Goldman2014}. 
In the following we treat $\gamma$ and $\kappa$ as two continuous parameters and we believe by changing the lattice wave vectors 
a wide parameter regime can be accessed in experiment \cite{Radic2012}.

Before ending this section, let us discuss the symmetries of the Hamiltonian.
The Hamiltonian is invariant under translational symmetry. There is no 
electron-hole symmetry and hence the chemical potential has to be adjusted during the DMFT loop. We find that 
unless the system is in a metallic phase or close to a critical point, the chemical potential $\mu=U/2$ 
leads to a density of one fermion per lattice site. At the spin-orbit coupling $\kappa=0$,  
one has $\bs{T}_{\hat{\bs y}}=\bs{\mathds{1}}$ and $\bs{T}_{\hat{\bs x}}=\bs{T}_{\hat{\bs x}+\hat{\bs y}}$, 
and hence the Hamiltonian is invariant under reflection with respect to the axis $(2\hat{\bs x}+\hat{\bs y})$, 
i.e., the horizontal dashed line in Fig. \ref{fig:hamiltonian}. In addition, the Hamiltonian is invariant 
under reflection with respect to the $\hat{\bs y}$ axis, provided that we apply the transformation $\gamma \longrightarrow -\gamma$ 
which maps $\bs{T}_{\hat{\bs x}}^\vpdag \longrightarrow \bs{T}_{\hat{\bs x}}^\dagg$. This allows one to limit 
$0\leq \gamma < 0.5$ at $\kappa=0$. The time-reversal operator is $\Theta=e^{-i\pi J_y}K$ where $J_y$ is the $\hat{\bs y}$-component 
of the spin operator and $K$ is the complex conjugate operator and we set $\hbar=1$. Under the time-reversal transformation the 
$SU(3)$ creation field operators transform as
\bearr
\Theta \Psi_{\bs r}^\dagg \Theta^{-1} 
= \left( c^\dagg_{{\bs r},-1}, -c^\dagg_{{\bs r}, 0},c^\dagg_{{\bs r}, +1} \right)&=:&\Psi_{\bs r}^\dagg(1-2\bs{J}_y^2) \nn \\
&=:& \Psi_{\bs r}^\dagg \bs{\Theta} K
\eearr
where $\bs{J}_y$ and $\bs{\Theta}$ are the matrix representations of the operators $J_y$ and $\Theta$. Since the Gell-Mann 
matrix $\bs{\lambda}_5$ is even under time-reversal, i.e., $\bs{\Theta} \bs{\lambda}_5 \bs{\Theta}^{-1}=+\bs{\lambda}_5$,
the Hamiltonian \eqref{eq:hamiltonian} is not time-reversal-invariant for any finite $\gamma$ and $\kappa$.

\section{metal-insulator transition} 
The $SU(3)$ Hubbard model at and near $1/2$-filling with anisotropic interactions has already been studied 
in detail in Refs. \onlinecite{Miyatake2010,Inaba2010,Inaba2012} where Fermi-liquid, superfluid, 
paired Mott insulator, and color-selective Mott insulator phases are characterized. At $1/3$-filling 
and in the large-$U$ limit, the fermionic $SU(3)$ Hubbard model can be effectively described  by the 
$SU(3)$ Heisenberg model, which is shown to have 3-sublattice magnetic order on both square 
and triangular lattices \cite{Toth2010,Bauer2012}. Thermal fluctuations destabilize this 
3-sublattice order into 2-sublattice order and subsequently into a paramagnetic 
phase \cite{Sotnikov2014,Sotnikov2015}.

Nevertheless, the interaction-driven transition between the metal and the 3-sublattice order Mott insulator 
phase at $1/3$-filling has not been addressed yet. 
In this section, we set the spin-orbit coupling $\gamma=\kappa=0$ which reduces the Hamiltonian 
\eqref{eq:hamiltonian} to the fermionic $SU(3)$ Hubbard model on the triangular lattice.
We fix the inverse temperature to $\beta t=20$.
The ED impurity solver is used with 4 bath sites and the results are checked versus 5 bath sites. 
We consider $30\times 30$ lattices with periodic boundary conditions. The results remain unchanged upon 
increasing the system size to $51\times 51$.

\begin{figure}[t]
    \centering
     \includegraphics[width=.93\linewidth,angle=0]{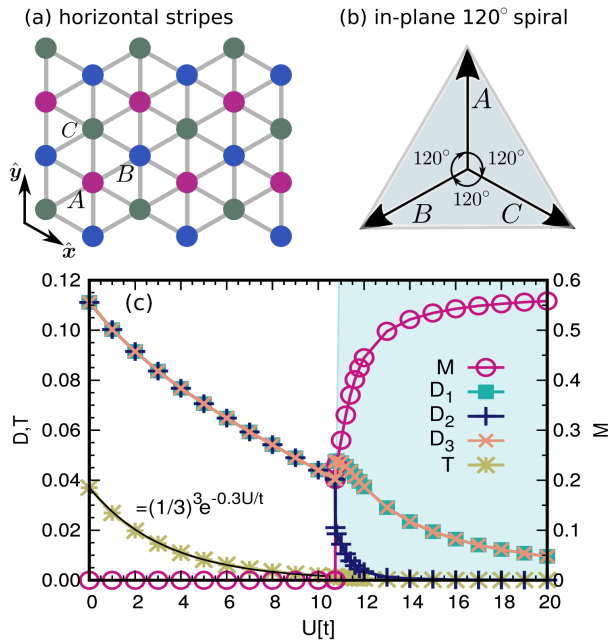}
     \caption{(color online). The 3-sublattice order (a) and the in-plane $120^\circ$ pseudospin order (b) 
     specifying the Mott insulator phase of the $SU(3)$ Hubbard model on the triangular lattice. (c)
     the three different double occupancies $D_1$, $D_2$, and $D_3$, the triple occupancy $T$, and 
     the local magnetization $M$ versus the Hubbard interaction $U$ in the $SU(3)$ Hubbard model.}
     \label{fig:mit}
\end{figure}

We identify a phase transition from a metallic  
to a Mott insulator phase  at $U_c\simeq 10.7t$ upon increasing the Hubbard interaction. The Mott 
insulator phase is characterized by the 3-sublattice order shown in Fig. \ref{fig:mit}a. The system consists of horizontal stripes 
with a sequence of, e.g., $A$, $B$, and $C$ stripes. We call this phase horizontal-stripe phase to be distinguished 
from more elaborate lattice patterns that we find in the next section. To reveal the 
pseudospin order we calculate the $8$-dimensional pseudospin vector 
\be 
\boldsymbol{\mathcal{S}}_{\bs r}^\vpdag := \frac{1}{2} \langle \Psi_{\boldsymbol r}^\dagg {\boldsymbol \lambda} \Psi_{\bs r}^\vpdag \rangle
\ee
where $\boldsymbol{\lambda}:=\left(\bs{\lambda}_1,\bs{\lambda}_2,\cdots,\bs{\lambda}_8 \right)$ is a vector made of 
Gell-Mann matrices. 
The angle between pseudospin vectors is computed using the scalar 
product. 
The Mott insulator phase shows an in-plane $120^\circ$ pseudospin order sketched in Fig. \ref{fig:mit}b. 
This is similar to the spiral long-range order in the $SU(2)$ Hubbard model on the triangular lattice. 
Due to the spontaneous breaking  of the $SU(3)$ symmetry, the plane in which the pseudospin vectors lie 
is not unique. With our unbiased real-space DMFT method we have been able to generate different of these 
degenerate states. The trivial solution is the one 
where $\langle c^\dagg_{\bs{r}\alpha} c^\vpdag_{\bs{r}\beta} \rangle = 0$ for any $\alpha \neq \beta$ and consequently 
the pseudospin vectors lie 
in the $\hat{\mathcal{S}}_3-\hat{\mathcal{S}}_8$ plane, where $\hat{\mathcal{S}}_i$ stands 
for the unit vector in the $i$th direction in the pseudospin space. In this state, at each 
sublattice one of the flavors has the dominant density 
and the densities of the other two flavors are equal. Similar spiral orders for frustrated classical spin  systems 
are recently created and detected in optical lattices \cite{Struck2011}.

There are three different double occupancies $D_1$, $D_2$, and $D_3$ corresponding to different  $\alpha$ and $\beta$ 
in $\langle n_{\bs{r}\alpha}^\vpdag n_{\bs{r}\beta}^\vpdag \rangle$. 
In Fig. \ref{fig:mit}c, we have plotted the double occupancies, 
the triple occupancy $T:=\langle n_{\bs{r},+1}^\vpdag n_{\bs{r},0}^\vpdag n_{\bs{r},-1}^\vpdag \rangle$ 
as well as the local magnetization 
$M:=|\boldsymbol{\mathcal{S}}_{\bs r}^\vpdag|$ versus the Hubbard interaction $U$. The three double occupancies are 
equal in the metallic phase and decrease upon increasing $U$ up to the transition point $U_c$. In the Mott phase, 
two of the double occupancies are equal and larger than the third one. Right above $U_c$, 
two of the double occupancies increase and the third one decreases sharply. At large values of $U$, the larger 
double occupancies decrease as power-law and the third one is negligible. The triple occupancy decreases exponentially 
versus $U$ and in the entire metallic phase can be fitted almost perfectly with the function $\langle n_{{\bs r}\alpha}^\vpdag \rangle^3 e^{-0.3U/t}$. 
The local magnetization $M$, similar to the double occupancies, 
changes sharply across the transition point and 
in the large-$U$ limit approaches the fully polarized value $1/\sqrt{3}\simeq 0.58$. 
We believe the data support a second 
order or very weakly first order transition. 

\section{spin-orbit-coupled Mott insulators}
We fix the Hubbard interaction to $U=15t$ and explore the effect of the spin-orbit 
coupling on the 3-sublattice magnetic order with horizontal stripes discussed in the previous section. 
The inverse temperature is fixed to $\beta t=20$. 
We have considered the ED impurity solver with 3 bath sites and checked the results versus 4 bath sites. 
We have mainly considered $30\times 30$ lattices, periodic boundary conditions, 
and unit cells as large as $6\times 3$. 
But we have checked for some selected points that the results remain unchanged upon 
increasing the system size to $48\times 48$ and increasing the unit cell to $12\times 6$. 
This allows us to find commensurate magnetic orders with relatively large periodicity.

\begin{figure*}[t]
    \centering
     \includegraphics[width=0.95\linewidth,angle=0]{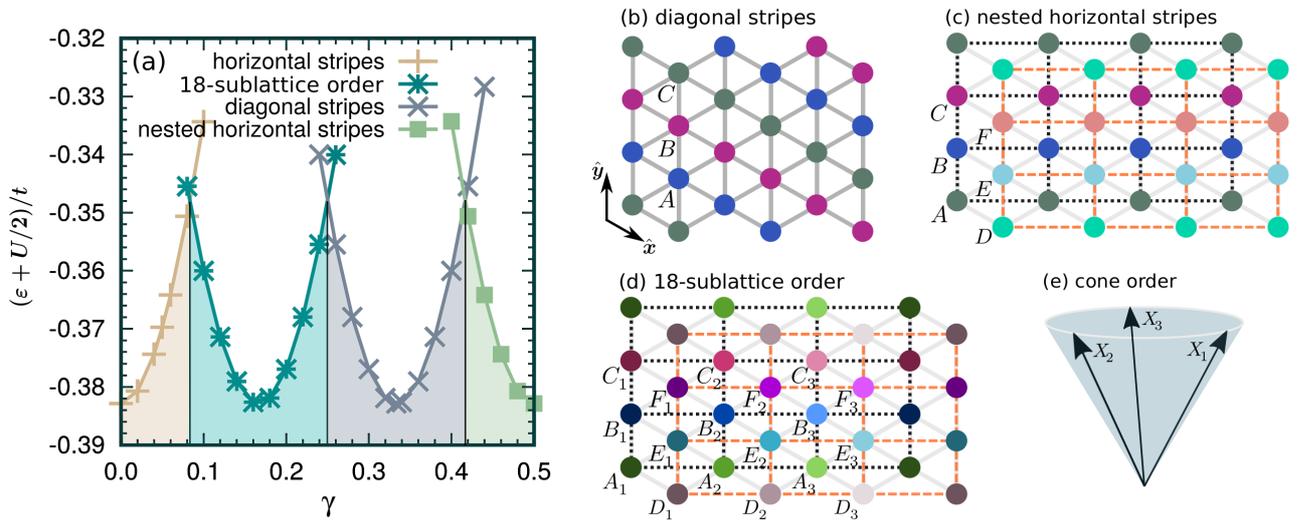}
     \caption{(color online). (a) Ground state energy per lattice site $\epsilon$ shifted by $U/2$ versus the spin-orbit 
     coupling $\gamma$ at $\kappa=0$ and $U=15t$. Schematic representation of diagonal stripes (b), nested horizontal 
     stripes (c), and $18$-sublattice order (d). (e) schematic sketch of the 3-site cone order.}
     \label{fig:mott1}
\end{figure*}

In contrast to the $SU(2)$ systems where spins are 3-dimensional objects, the $8$-dimensional nature of 
the pseudospin vectors in the $SU(3)$ systems make it difficult to identify and name the pseudospin order. 
We have mostly concentrated on the stabilization of different lattice patterns induced by the spin-orbit 
coupling. The pseudospin order is discussed only in some cases.
For any values of $\gamma$ and $\kappa$, we always find that the norm of the pseudospin vector is the same on 
every lattice site.

We first set $\kappa=0$ and study the effect of $\gamma$. 
In Fig. \ref{fig:mott1}, we have plotted the ground state energy per lattice site $\epsilon$ shifted by $U/2$ 
versus $\gamma$, panel (a), as well as the schematic representation of different lattice patterns which appear, panels (b)-(d). 
We always find the pseudospin vector  
in the five dimensional ${\hat{\mathcal{S}}_1-\hat{\mathcal{S}}_3-\hat{\mathcal{S}}_4-\hat{\mathcal{S}}_6-\hat{\mathcal{S}}_8}$ 
space, i.e., $\langle \mathcal{S}_2 \rangle=\langle \mathcal{S}_5 \rangle=\langle \mathcal{S}_7 \rangle=0$. 
However, we do not exclude the possibility that there might be other degenerate solutions as the $SU(3)$ symmetry 
is not fully broken by $\gamma \neq 0$. Moreover, we find that the pseudospin vectors in the $\hat{\bs y}$ direction always 
form an in-plane $120^\circ$ order as depicted in Fig. \ref{fig:mit}b.

We discuss the simpler phases in Fig. \ref{fig:mott1}a first. One can see from Fig. \ref{fig:mott1}a 
that the ground state energy of the phase with horizontal stripes increases upon increasing $\gamma$. 
This phase is stable up to $\gamma\simeq 1/12$ where a first order transition takes place. 
For $1/4 \lesssim \gamma \lesssim 5/12$, the system has a ferromagnetic order along the $\hat{\bs x}$ direction 
and an in-plane $120^\circ$ spiral order along the $\hat{\bs y}$ direction. This phase is schematically displayed 
in Fig. \ref{fig:mott1}b. 
The lattice comprises diagonal stripes with a sequence of a 3-color cycle. 
Due to the spontaneous breaking of the horizontal reflection symmetry, 
this state is degenerate with the one in which diagonal 
stripes are oriented along $\hat{\bs x}+\hat{\bs y}$ direction. 
We have explicitly checked this degeneracy. As the hopping matrices at $\gamma=1/3$ and $\kappa=0$  simplify to
\be 
\bs{T}_{\hat{\bs x}}=\bs{T}_{\hat{\bs x}+\hat{\bs y}}
=
\bpm
0 & 1 & 0 \\
0 & 0 & 1 \\
1 & 0 & 0
\epm, \quad
\bs{T}_{\hat{\bs y}}=\hat{\mathds{1}} \quad,
\ee
a diagonal-stripe phase in the vicinity of $\gamma=1/3$ is what one would expect from 
a second order perturbation theory in the large-$U$ limit \cite{Duan2003}. The $SU(2)$ counterpart of such a phase 
would have the ferromagnetic and the antiferromagnetic orders along 
$\hat{\bs x}$ and $\hat{\bs y}$ directions and is found, for example, as a result of a Rashba-like 
spin-orbit coupling in both bosonic \cite{Radic2012} and fermionic systems \cite{Cocks2012,Orth2013}. 
This is usually called collinear or nematic order.

For $5/12 \lesssim \gamma \lesssim 1/2$, we identify a phase with 6-sublattice order  
shown schematically in Fig. \ref{fig:mott1}c. 
In contrast to the 3-sublattice order sketched in Fig. \ref{fig:mit}a, this is a phase 
with the length of the unit cell along $\hat{\bs x}$ direction doubled.
One can 
understand this phase as two penetrating rectangular lattices specified by 
black dotted lines and by orange dashed lines in Fig. \ref{fig:mott1}c. 
On each rectangular lattice there is a ferromagnetic order along the horizontal direction 
and an in-plane $120^\circ$ spiral order along the vertical direction,
i.e., the pseudospin 
vectors at the sites $A$, $B$, and $C$ obey the relation
$\bs{\mathcal{S}}_{\! A}+\bs{\mathcal{S}}_{\! B}+\bs{\mathcal{S}}_{\! C}=\bs{0}$ and likewise 
for the pseudospin vectors at the sites $D$, $E$, and $F$. 
One notices that the norm of the pseudospin vector is the same on every lattice site.
Each rectangular 
lattice represents horizontal stripes with a sequence of a 3-color cycle. Although the 
states over the two rectangular lattices are obviously coupled, we have not been able 
to find a specific relation between the pseudospin order at the sites  $A$, $B$, and $C$ and 
the pseudospin order at the sites $D$, $E$, and $F$.
One can interpret this  state also over the full triangular lattice as consisted of horizontal stripes 
with a succession of a 6-color cycle. 
We refer to this state as nested horizontal stripes. 

We proceed with the state which appears for $1/12 \lesssim \gamma \lesssim 1/4$ in Fig. \ref{fig:mott1}a. 
The symmetry in this phase is quite reduced and the system shows an 18-sublattice order. 
This phase is schematically shown in Fig. \ref{fig:mott1}d with a 6-site periodicity 
along $\hat{\bs x}$ and a 3-site periodicity along $\hat{\bs y}$ direction. 
To have a simpler understanding 
of this state, we have interpreted the triangular lattice again as two penetrating rectangular 
lattices specified by black dotted lines and orange dashed lines in Fig. \ref{fig:mott1}d. 
The letters $A$, $B$, and $C$ and the letters $D$, $E$, and $F$ distinguish the sites 
of the two rectangular lattices. On each rectangular lattice, one has a 3-site cone order 
along the horizontal direction shown in Fig. \ref{fig:mott1}e, where $X$ stands for any of 
the letters from $A$ to $F$. The $SU(3)$ cone order is characterized by equal angles between any  
pair of the pseudospin vectors, 
but the pseudospin vectors 
are not coplanar  like the spiral order in Fig. \ref{fig:mit}b.
The state remains invariant under translation along $2\hat{\bs x}+\hat{\bs y}$ direction 
provided that one applies the following two independent transformations in the pseudospin space: 
(I) a clockwise $120^\circ$ rotation in $\hat{\bs{\mathcal{S}}}_3-\hat{\bs{\mathcal{S}}}_8$ 
space and (II) the cyclic permutation 
$\hat{\bs{\mathcal{S}}}_1 \rightarrow \hat{\bs{\mathcal{S}}}_4 \rightarrow 
\hat{\bs{\mathcal{S}}}_6 \rightarrow \hat{\bs{\mathcal{S}}}_1$.
The state with the opposite chirality appears in the area $-1/4 \lesssim \gamma \lesssim -1/12$ due 
to the vertical reflection symmetry. We have checked this fact explicitly.
Similar to all the previous phases that we discussed, the system shows an in-plane 
$120^\circ$ spiral order in the $\hat{\bs y}$ direction.

We notice the coexistence regions near the transition points in Fig. \ref{fig:mott1}a, which 
indicates the requirement for the calculation of the ground state energy in order to determine 
the precise positions of the transition points. The transition points occurring at the nice fractional numbers 
$\gamma \simeq 1/12$, $1/4$, and $5/12$ is reminiscent of the classical analysis of spin models. This 
is plausible because in the large-$U$ limit the local fluctuations are frozen and our unbiased 
real-space DMFT becomes equivalent to the classical approximation. 
The minima of the energy in Fig. \ref{fig:mott1}a occurring at $\gamma \simeq 0$, $1/6$, $1/3$, and $1/2$ 
are interestingly very close, 
which illustrates how different long-range orders become favorable in different spin-orbit coupling regimes. 

We turn now to the case of finite $\kappa$. The Hamiltonian is invariant under 
the transformation $\kappa \rightarrow \kappa +2$.
In addition, we find that the phase diagram is symmetric with respect to $\kappa=0$. Hence, we restrict $0 \leq \kappa \leq 1$.
Upon introducing a finite value of $\kappa$, all the 8 components of the pseudospin vector become finite, making 
the recognition and the discussion of the pseudospin order more complicated. In the 
following we focus only on ordering patterns in real space. 

\begin{figure}[t]
    \centering
     \includegraphics[width=0.99\linewidth,angle=0]{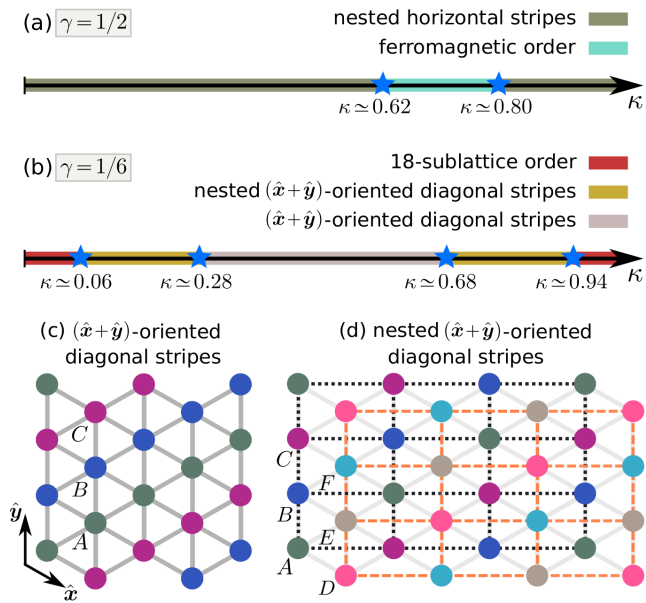}
     \caption{(color online). Schematic phase diagram of the Hamiltonian \eqref{eq:hamiltonian} 
     versus $\kappa$ at $\gamma=1/2$ (a) and $\gamma=1/6$ (b) for the Hubbard interaction $U=15t$. 
     Schematic representation of diagonal stripes (c) and nested diagonal stripes (d) with 
     $\hat{\bs x}+\hat{\bs y}$ orientation.}
     \label{fig:mott2}
\end{figure}

We consider the values $\gamma =0$, $1/6$, $1/3$, and $1/2$ and study the effect of $\kappa$ on the 
different lattice patterns that we discussed. We find that the horizontal-stripe pattern in Fig. \ref{fig:mit}a 
as well as the diagonal-stripe pattern in Fig. \ref{fig:mott1}b remain stable 
upon introducing the spin-orbit coupling $\kappa$. 
The phase diagram versus $\kappa$ at $\gamma=1/2$ is presented in Fig. \ref{fig:mott2}a. 
At $\gamma=1/2$, an intermediate ferromagnetic phase 
appears for $0.62 \lesssim \kappa \lesssim 0.80$. Below and above this region 
we find the nested horizontal-stripe pattern sketched 
in Fig. \ref{fig:mott1}c. 
The phase digram versus $\kappa$ at $\gamma=1/6$ is plotted in Fig. \ref{fig:mott2}b. 
In addition to the 18-sublattice order, see Fig. \ref{fig:mott1}d, near $\kappa=0$ and $\kappa=1$, we detect   
a  diagonal-stripe order and a nested diagonal-stripe order. In both phases, 
stripes are oriented in the $\hat{\bs x}+\hat{\bs y}$ direction 
and they are depicted in Figs. \ref{fig:mott2}c 
and \ref{fig:mott2}d. In the nested diagonal-stripe state, the triangular lattice is seen 
as two penetrating rectangular lattices, each one forming a $(\hat{\bs x}+\hat{\bs y})$-oriented diagonal-stripe order. 
The phase diagram in Fig. \ref{fig:mott2}b is approximately symmetric with respect to $\kappa=1/2$, but this does not apply 
to physical quantities such as, for example, the ground state energy. Upon increasing $\kappa$ from $0$ to $1/2$ at $\gamma=1/6$ the symmetry 
of the lattice is restored; first from the 18-sublattice order to the 6-sublattice order with nested diagonal stripes and 
subsequently to the 3-sublattice order with diagonal stripes. 

\section{Summary and Outlook}

In recent years, there has been a noticeable development on realization of artificial gauge fields 
in optical lattices, 
which led to the implementation of the fundamental models 
such as the Haldane \cite{Jotzu2014} and Hofstadter Hamiltonians \cite{Aidelsburger2013,Aidelsburger2014,Miyake2013}. 
Beside experimental achievments, there has been 
a number of experimental proposals and theoretical predictions especially for non-Abelian gauge fields and 
effective spin-orbit coupling in $SU(2)$ and $SU(3)$ systems \cite{Goldman2014}. While the spin-orbit coupling in 
non-interacting systems is essential to realize topological bands \cite{Hasan2010}, 
in strongly interacting regimes it can stabilize Mott insulators 
with exotic long-range orders \cite{Cole2012,Hickey2015,Cocks2012,Radic2012,Orth2013}. 

In this work, we have investigated the fermionic $SU(3)$ Hubbard model in the presence of spin-orbit 
coupling on the triangular lattice. The $SU(3)$ Hubbard model shows a transition from a metallic phase 
to a Mott insulator \cite{DelRe2017}, which we have studied in details. The Mott phase has in-plane $120^\circ$ 
spiral pseudospin order. The spin-orbit coupling drives this 3-sublattice Mott insulator to Mott states 
with various types of lattice patterns such as horizontal stripes and diagonal stripes with different 
orientations. In addition, we find more complex lattice orders which we have interpreted as two nested 
rectangular lattices with horizontal and diagonal stripes each. Due to the complex 8-dimensional 
nature of the pseudospin vector, the pseudospin order is discussed only in some cases where ferromagnetic, spiral, 
and cone orders are recognized. While this work is mainly devoted to the spin-orbit-coupled Mott 
insulators, we have shown that the Hamiltonian considered here has interesting topological features through 
the connection with the $SU(3)$ Hatsugai-Harper-Hofstadter-Hubbard model. 
The real-space DMFT method equipped with the continuous-time quantum-Monte-Carlo 
impurity solver enables us to address edge states of interacting topological phases on cylindrical geometries. 
An alternative approach to discuss topological phase transitions would be the calculation of 
the Chern number, which   is 
already formulated for $SU(3)$ systems in the non-interacting case \cite{Barnett2012}.
This could be extended to interacting phases using, for example, the effective topological 
Hamiltonian approach \cite{Wang2012}.
However, fractional topological insulators still remain out of reach for (real-space) DMFT calculations 
based on a local self-energy.

Having implemented a real-space DMFT which 
can work for $SU(N)$ systems with arbitrary $N$, it 
would not be difficult to generalize it to the case of cluster real-space DMFT.
Strictly speaking, one requires in the Anderson impurity model \eqref{eq:aim} to consider a 
chemical potential matrix $\bs{\mu}_{\bs{r}}$ with finite off-diagonal elements and to make the bath onsite 
energies $\varepsilon_l^{\bs{r}}$ flavor-dependent.
The main computational restriction would be the diagonalization of the Anderson impurity model. One would need 
to focus on zero temperature properties and use the Lanczos algorithm to reach a larger number of bath sites.
By taking into account non-local quantum fluctuations this cluster real-space DMFT would allow us to identify 
phases such as quantum spin liquids \cite{Hermele2009}, valence-bond solid states  
with different dimerization patterns \cite{Hermele2011,Honerkamp2004},  and 
orientational bond states with and without magnetic long-range order \cite{Torbati2016}.

While our system describes the single-orbital Hubbard model characterized by a single Hubbard $U$, 
two-orbital $SU(N)$ Mott insulators involving intra-orbital Hubbard term and inter-orbital direct 
and exchange interactions can also be realized in optical lattices \cite{Gorshkov2010a}.
It is left for future research to explore the topological properties of the model at finite 
spin-orbit coupling and different interaction strengths, to consider multi-orbital $SU(N)$ systems, 
and to study the effect of non-local 
quantum fluctuations by going beyond a local self-energy.

\section*{acknowledgment}
We would like to thank Nathan Goldman, Bernhard Irsigler, Jaromir Panas, Andrii Sotnikov, Christof Weitenberg, 
and Jun-Hui Zheng for useful discussions.  
This research was funded by the Deutsche Forschungsgemeinschaft
(DFG, German Research Foundation) via Research Unit FOR 2414
under project number 277974659. This work was also supported by 
the high performance computing center LOEWE-CSC.

\section*{References}

\begin{thebibliography}{54}%
\makeatletter
\providecommand \@ifxundefined [1]{%
 \@ifx{#1\undefined}
}%
\providecommand \@ifnum [1]{%
 \ifnum #1\expandafter \@firstoftwo
 \else \expandafter \@secondoftwo
 \fi
}%
\providecommand \@ifx [1]{%
 \ifx #1\expandafter \@firstoftwo
 \else \expandafter \@secondoftwo
 \fi
}%
\providecommand \natexlab [1]{#1}%
\providecommand \enquote  [1]{``#1''}%
\providecommand \bibnamefont  [1]{#1}%
\providecommand \bibfnamefont [1]{#1}%
\providecommand \citenamefont [1]{#1}%
\providecommand \href@noop [0]{\@secondoftwo}%
\providecommand \href [0]{\begingroup \@sanitize@url \@href}%
\providecommand \@href[1]{\@@startlink{#1}\@@href}%
\providecommand \@@href[1]{\endgroup#1\@@endlink}%
\providecommand \@sanitize@url [0]{\catcode `\\12\catcode `\$12\catcode
  `\&12\catcode `\#12\catcode `\^12\catcode `\_12\catcode `\%12\relax}%
\providecommand \@@startlink[1]{}%
\providecommand \@@endlink[0]{}%
\providecommand \url  [0]{\begingroup\@sanitize@url \@url }%
\providecommand \@url [1]{\endgroup\@href {#1}{\urlprefix }}%
\providecommand \urlprefix  [0]{URL }%
\providecommand \Eprint [0]{\href }%
\providecommand \doibase [0]{http://dx.doi.org/}%
\providecommand \selectlanguage [0]{\@gobble}%
\providecommand \bibinfo  [0]{\@secondoftwo}%
\providecommand \bibfield  [0]{\@secondoftwo}%
\providecommand \translation [1]{[#1]}%
\providecommand \BibitemOpen [0]{}%
\providecommand \bibitemStop [0]{}%
\providecommand \bibitemNoStop [0]{.\EOS\space}%
\providecommand \EOS [0]{\spacefactor3000\relax}%
\providecommand \BibitemShut  [1]{\csname bibitem#1\endcsname}%
\let\auto@bib@innerbib\@empty
\bibitem [{\citenamefont {Anderson}\ \emph {et~al.}(1995)\citenamefont
  {Anderson}, \citenamefont {Ensher}, \citenamefont {Matthews}, \citenamefont
  {Wieman},\ and\ \citenamefont {Cornell}}]{Anderson1995}%
  \BibitemOpen
  \bibfield  {author} {\bibinfo {author} {\bibfnamefont {M.~H.}\ \bibnamefont
  {Anderson}}, \bibinfo {author} {\bibfnamefont {J.~R.}\ \bibnamefont
  {Ensher}}, \bibinfo {author} {\bibfnamefont {M.~R.}\ \bibnamefont
  {Matthews}}, \bibinfo {author} {\bibfnamefont {C.~E.}\ \bibnamefont
  {Wieman}}, \ and\ \bibinfo {author} {\bibfnamefont {E.~A.}\ \bibnamefont
  {Cornell}},\ }\href {\doibase 10.1126/science.269.5221.198} {\bibfield
  {journal} {\bibinfo  {journal} {Science}\ }\textbf {\bibinfo {volume}
  {269}},\ \bibinfo {pages} {198} (\bibinfo {year} {1995})}\BibitemShut
  {NoStop}%
\bibitem [{\citenamefont {Bloch}(2005)}]{Bloch2005}%
  \BibitemOpen
  \bibfield  {author} {\bibinfo {author} {\bibfnamefont {I.}~\bibnamefont
  {Bloch}},\ }\href {http://dx.doi.org/10.1038/nphys138} {\bibfield  {journal}
  {\bibinfo  {journal} {Nature Physics}\ }\textbf {\bibinfo {volume} {1}},\
  \bibinfo {pages} {23} (\bibinfo {year} {2005})}\BibitemShut {NoStop}%
\bibitem [{\citenamefont {Georgescu}\ \emph {et~al.}(2014)\citenamefont
  {Georgescu}, \citenamefont {Ashhab},\ and\ \citenamefont
  {Nori}}]{Georgescu2014}%
  \BibitemOpen
  \bibfield  {author} {\bibinfo {author} {\bibfnamefont {I.~M.}\ \bibnamefont
  {Georgescu}}, \bibinfo {author} {\bibfnamefont {S.}~\bibnamefont {Ashhab}}, \
  and\ \bibinfo {author} {\bibfnamefont {F.}~\bibnamefont {Nori}},\ }\href
  {\doibase 10.1103/RevModPhys.86.153} {\bibfield  {journal} {\bibinfo
  {journal} {Rev. Mod. Phys.}\ }\textbf {\bibinfo {volume} {86}},\ \bibinfo
  {pages} {153} (\bibinfo {year} {2014})}\BibitemShut {NoStop}%
\bibitem [{\citenamefont {Haldane}(1988)}]{Haldane1988}%
  \BibitemOpen
  \bibfield  {author} {\bibinfo {author} {\bibfnamefont {F.~D.~M.}\
  \bibnamefont {Haldane}},\ }\href {\doibase 10.1103/PhysRevLett.61.2015}
  {\bibfield  {journal} {\bibinfo  {journal} {Phys. Rev. Lett.}\ }\textbf
  {\bibinfo {volume} {61}},\ \bibinfo {pages} {2015} (\bibinfo {year}
  {1988})}\BibitemShut {NoStop}%
\bibitem [{\citenamefont {Jotzu}\ \emph {et~al.}(2014)\citenamefont {Jotzu},
  \citenamefont {Messer}, \citenamefont {Desbuquois}, \citenamefont {Lebrat},
  \citenamefont {Uehlinger}, \citenamefont {Greif},\ and\ \citenamefont
  {Esslinger}}]{Jotzu2014}%
  \BibitemOpen
  \bibfield  {author} {\bibinfo {author} {\bibfnamefont {G.}~\bibnamefont
  {Jotzu}}, \bibinfo {author} {\bibfnamefont {M.}~\bibnamefont {Messer}},
  \bibinfo {author} {\bibfnamefont {R.}~\bibnamefont {Desbuquois}}, \bibinfo
  {author} {\bibfnamefont {M.}~\bibnamefont {Lebrat}}, \bibinfo {author}
  {\bibfnamefont {T.}~\bibnamefont {Uehlinger}}, \bibinfo {author}
  {\bibfnamefont {D.}~\bibnamefont {Greif}}, \ and\ \bibinfo {author}
  {\bibfnamefont {T.}~\bibnamefont {Esslinger}},\ }\href
  {http://dx.doi.org/10.1038/nature13915} {\bibfield  {journal} {\bibinfo
  {journal} {Nature}\ }\textbf {\bibinfo {volume} {515}},\ \bibinfo {pages}
  {237} (\bibinfo {year} {2014})}\BibitemShut {NoStop}%
\bibitem [{\citenamefont {Aidelsburger}\ \emph {et~al.}(2011)\citenamefont
  {Aidelsburger}, \citenamefont {Atala}, \citenamefont {Nascimb\`ene},
  \citenamefont {Trotzky}, \citenamefont {Chen},\ and\ \citenamefont
  {Bloch}}]{Aidelsburger2011}%
  \BibitemOpen
  \bibfield  {author} {\bibinfo {author} {\bibfnamefont {M.}~\bibnamefont
  {Aidelsburger}}, \bibinfo {author} {\bibfnamefont {M.}~\bibnamefont {Atala}},
  \bibinfo {author} {\bibfnamefont {S.}~\bibnamefont {Nascimb\`ene}}, \bibinfo
  {author} {\bibfnamefont {S.}~\bibnamefont {Trotzky}}, \bibinfo {author}
  {\bibfnamefont {Y.-A.}\ \bibnamefont {Chen}}, \ and\ \bibinfo {author}
  {\bibfnamefont {I.}~\bibnamefont {Bloch}},\ }\href {\doibase
  10.1103/PhysRevLett.107.255301} {\bibfield  {journal} {\bibinfo  {journal}
  {Phys. Rev. Lett.}\ }\textbf {\bibinfo {volume} {107}},\ \bibinfo {pages}
  {255301} (\bibinfo {year} {2011})}\BibitemShut {NoStop}%
\bibitem [{\citenamefont {Aidelsburger}\ \emph {et~al.}(2013)\citenamefont
  {Aidelsburger}, \citenamefont {Atala}, \citenamefont {Lohse}, \citenamefont
  {Barreiro}, \citenamefont {Paredes},\ and\ \citenamefont
  {Bloch}}]{Aidelsburger2013}%
  \BibitemOpen
  \bibfield  {author} {\bibinfo {author} {\bibfnamefont {M.}~\bibnamefont
  {Aidelsburger}}, \bibinfo {author} {\bibfnamefont {M.}~\bibnamefont {Atala}},
  \bibinfo {author} {\bibfnamefont {M.}~\bibnamefont {Lohse}}, \bibinfo
  {author} {\bibfnamefont {J.~T.}\ \bibnamefont {Barreiro}}, \bibinfo {author}
  {\bibfnamefont {B.}~\bibnamefont {Paredes}}, \ and\ \bibinfo {author}
  {\bibfnamefont {I.}~\bibnamefont {Bloch}},\ }\href {\doibase
  10.1103/PhysRevLett.111.185301} {\bibfield  {journal} {\bibinfo  {journal}
  {Phys. Rev. Lett.}\ }\textbf {\bibinfo {volume} {111}},\ \bibinfo {pages}
  {185301} (\bibinfo {year} {2013})}\BibitemShut {NoStop}%
\bibitem [{\citenamefont {Miyake}\ \emph {et~al.}(2013)\citenamefont {Miyake},
  \citenamefont {Siviloglou}, \citenamefont {Kennedy}, \citenamefont {Burton},\
  and\ \citenamefont {Ketterle}}]{Miyake2013}%
  \BibitemOpen
  \bibfield  {author} {\bibinfo {author} {\bibfnamefont {H.}~\bibnamefont
  {Miyake}}, \bibinfo {author} {\bibfnamefont {G.~A.}\ \bibnamefont
  {Siviloglou}}, \bibinfo {author} {\bibfnamefont {C.~J.}\ \bibnamefont
  {Kennedy}}, \bibinfo {author} {\bibfnamefont {W.~C.}\ \bibnamefont {Burton}},
  \ and\ \bibinfo {author} {\bibfnamefont {W.}~\bibnamefont {Ketterle}},\
  }\href {\doibase 10.1103/PhysRevLett.111.185302} {\bibfield  {journal}
  {\bibinfo  {journal} {Phys. Rev. Lett.}\ }\textbf {\bibinfo {volume} {111}},\
  \bibinfo {pages} {185302} (\bibinfo {year} {2013})}\BibitemShut {NoStop}%
\bibitem [{\citenamefont {Aidelsburger}\ \emph {et~al.}(2014)\citenamefont
  {Aidelsburger}, \citenamefont {Lohse}, \citenamefont {Schweizer},
  \citenamefont {Atala}, \citenamefont {Barreiro}, \citenamefont
  {NascimbÃ¨ne}, \citenamefont {Cooper}, \citenamefont {Bloch},\ and\
  \citenamefont {Goldman}}]{Aidelsburger2014}%
  \BibitemOpen
  \bibfield  {author} {\bibinfo {author} {\bibfnamefont {M.}~\bibnamefont
  {Aidelsburger}}, \bibinfo {author} {\bibfnamefont {M.}~\bibnamefont {Lohse}},
  \bibinfo {author} {\bibfnamefont {C.}~\bibnamefont {Schweizer}}, \bibinfo
  {author} {\bibfnamefont {M.}~\bibnamefont {Atala}}, \bibinfo {author}
  {\bibfnamefont {J.}~\bibnamefont {Barreiro}}, \bibinfo {author}
  {\bibfnamefont {S.}~\bibnamefont {NascimbÃ¨ne}}, \bibinfo {author}
  {\bibfnamefont {N.}~\bibnamefont {Cooper}}, \bibinfo {author} {\bibfnamefont
  {I.}~\bibnamefont {Bloch}}, \ and\ \bibinfo {author} {\bibfnamefont
  {N.}~\bibnamefont {Goldman}},\ }\href {http://dx.doi.org/10.1038/nphys3171}
  {\bibfield  {journal} {\bibinfo  {journal} {Nature Physics}\ }\textbf
  {\bibinfo {volume} {11}},\ \bibinfo {pages} {162} (\bibinfo {year}
  {2014})}\BibitemShut {NoStop}%
\bibitem [{\citenamefont {Fl\"{a}schner}\ \emph {et~al.}(2016)\citenamefont
  {Fl\"{a}schner}, \citenamefont {Rem}, \citenamefont {Tarnowski},
  \citenamefont {Vogel}, \citenamefont {L\"{u}hmann}, \citenamefont
  {Sengstock},\ and\ \citenamefont {Weitenberg}}]{Flaschner2016}%
  \BibitemOpen
  \bibfield  {author} {\bibinfo {author} {\bibfnamefont {N.}~\bibnamefont
  {Fl\"{a}schner}}, \bibinfo {author} {\bibfnamefont {B.~S.}\ \bibnamefont
  {Rem}}, \bibinfo {author} {\bibfnamefont {M.}~\bibnamefont {Tarnowski}},
  \bibinfo {author} {\bibfnamefont {D.}~\bibnamefont {Vogel}}, \bibinfo
  {author} {\bibfnamefont {D.-S.}\ \bibnamefont {L\"{u}hmann}}, \bibinfo
  {author} {\bibfnamefont {K.}~\bibnamefont {Sengstock}}, \ and\ \bibinfo
  {author} {\bibfnamefont {C.}~\bibnamefont {Weitenberg}},\ }\href
  {http://science.sciencemag.org/content/352/6289/1091.abstract} {\bibfield
  {journal} {\bibinfo  {journal} {Science}\ }\textbf {\bibinfo {volume}
  {352}},\ \bibinfo {pages} {1091} (\bibinfo {year} {2016})}\BibitemShut
  {NoStop}%
\bibitem [{\citenamefont {Jaksch}\ and\ \citenamefont
  {Zoller}(2003)}]{Jaksch2003}%
  \BibitemOpen
  \bibfield  {author} {\bibinfo {author} {\bibfnamefont {D.}~\bibnamefont
  {Jaksch}}\ and\ \bibinfo {author} {\bibfnamefont {P.}~\bibnamefont
  {Zoller}},\ }\href {http://stacks.iop.org/1367-2630/5/i=1/a=356} {\bibfield
  {journal} {\bibinfo  {journal} {New Journal of Physics}\ }\textbf {\bibinfo
  {volume} {5}},\ \bibinfo {pages} {56} (\bibinfo {year} {2003})}\BibitemShut
  {NoStop}%
\bibitem [{\citenamefont {Osterloh}\ \emph {et~al.}(2005)\citenamefont
  {Osterloh}, \citenamefont {Baig}, \citenamefont {Santos}, \citenamefont
  {Zoller},\ and\ \citenamefont {Lewenstein}}]{Osterloh2005}%
  \BibitemOpen
  \bibfield  {author} {\bibinfo {author} {\bibfnamefont {K.}~\bibnamefont
  {Osterloh}}, \bibinfo {author} {\bibfnamefont {M.}~\bibnamefont {Baig}},
  \bibinfo {author} {\bibfnamefont {L.}~\bibnamefont {Santos}}, \bibinfo
  {author} {\bibfnamefont {P.}~\bibnamefont {Zoller}}, \ and\ \bibinfo {author}
  {\bibfnamefont {M.}~\bibnamefont {Lewenstein}},\ }\href {\doibase
  10.1103/PhysRevLett.95.010403} {\bibfield  {journal} {\bibinfo  {journal}
  {Phys. Rev. Lett.}\ }\textbf {\bibinfo {volume} {95}},\ \bibinfo {pages}
  {010403} (\bibinfo {year} {2005})}\BibitemShut {NoStop}%
\bibitem [{\citenamefont {Satija}\ \emph {et~al.}(2006)\citenamefont {Satija},
  \citenamefont {Dakin},\ and\ \citenamefont {Clark}}]{Satija2006}%
  \BibitemOpen
  \bibfield  {author} {\bibinfo {author} {\bibfnamefont {I.~I.}\ \bibnamefont
  {Satija}}, \bibinfo {author} {\bibfnamefont {D.~C.}\ \bibnamefont {Dakin}}, \
  and\ \bibinfo {author} {\bibfnamefont {C.~W.}\ \bibnamefont {Clark}},\ }\href
  {\doibase 10.1103/PhysRevLett.97.216401} {\bibfield  {journal} {\bibinfo
  {journal} {Phys. Rev. Lett.}\ }\textbf {\bibinfo {volume} {97}},\ \bibinfo
  {pages} {216401} (\bibinfo {year} {2006})}\BibitemShut {NoStop}%
\bibitem [{\citenamefont {Goldman}\ \emph {et~al.}(2010)\citenamefont
  {Goldman}, \citenamefont {Satija}, \citenamefont {Nikolic}, \citenamefont
  {Bermudez}, \citenamefont {Martin-Delgado}, \citenamefont {Lewenstein},\ and\
  \citenamefont {Spielman}}]{Goldman2010}%
  \BibitemOpen
  \bibfield  {author} {\bibinfo {author} {\bibfnamefont {N.}~\bibnamefont
  {Goldman}}, \bibinfo {author} {\bibfnamefont {I.}~\bibnamefont {Satija}},
  \bibinfo {author} {\bibfnamefont {P.}~\bibnamefont {Nikolic}}, \bibinfo
  {author} {\bibfnamefont {A.}~\bibnamefont {Bermudez}}, \bibinfo {author}
  {\bibfnamefont {M.~A.}\ \bibnamefont {Martin-Delgado}}, \bibinfo {author}
  {\bibfnamefont {M.}~\bibnamefont {Lewenstein}}, \ and\ \bibinfo {author}
  {\bibfnamefont {I.~B.}\ \bibnamefont {Spielman}},\ }\href {\doibase
  10.1103/PhysRevLett.105.255302} {\bibfield  {journal} {\bibinfo  {journal}
  {Phys. Rev. Lett.}\ }\textbf {\bibinfo {volume} {105}},\ \bibinfo {pages}
  {255302} (\bibinfo {year} {2010})}\BibitemShut {NoStop}%
\bibitem [{\citenamefont {Wu}\ \emph {et~al.}(2003)\citenamefont {Wu},
  \citenamefont {Hu},\ and\ \citenamefont {Zhang}}]{Wu2003}%
  \BibitemOpen
  \bibfield  {author} {\bibinfo {author} {\bibfnamefont {C.}~\bibnamefont
  {Wu}}, \bibinfo {author} {\bibfnamefont {J.-p.}\ \bibnamefont {Hu}}, \ and\
  \bibinfo {author} {\bibfnamefont {S.-c.}\ \bibnamefont {Zhang}},\ }\href
  {\doibase 10.1103/PhysRevLett.91.186402} {\bibfield  {journal} {\bibinfo
  {journal} {Phys. Rev. Lett.}\ }\textbf {\bibinfo {volume} {91}},\ \bibinfo
  {pages} {186402} (\bibinfo {year} {2003})}\BibitemShut {NoStop}%
\bibitem [{\citenamefont {Honerkamp}\ and\ \citenamefont
  {Hofstetter}(2004)}]{Honerkamp2004}%
  \BibitemOpen
  \bibfield  {author} {\bibinfo {author} {\bibfnamefont {C.}~\bibnamefont
  {Honerkamp}}\ and\ \bibinfo {author} {\bibfnamefont {W.}~\bibnamefont
  {Hofstetter}},\ }\href {\doibase 10.1103/PhysRevLett.92.170403} {\bibfield
  {journal} {\bibinfo  {journal} {Phys. Rev. Lett.}\ }\textbf {\bibinfo
  {volume} {92}},\ \bibinfo {pages} {170403} (\bibinfo {year}
  {2004})}\BibitemShut {NoStop}%
\bibitem [{\citenamefont {Gorshkov}\ \emph {et~al.}(2010)\citenamefont
  {Gorshkov}, \citenamefont {Hermele}, \citenamefont {Gurarie}, \citenamefont
  {Xu}, \citenamefont {Julienne}, \citenamefont {Ye}, \citenamefont {Zoller},
  \citenamefont {Demler}, \citenamefont {Lukin},\ and\ \citenamefont
  {Rey}}]{Gorshkov2010a}%
  \BibitemOpen
  \bibfield  {author} {\bibinfo {author} {\bibfnamefont {A.~V.}\ \bibnamefont
  {Gorshkov}}, \bibinfo {author} {\bibfnamefont {M.}~\bibnamefont {Hermele}},
  \bibinfo {author} {\bibfnamefont {V.}~\bibnamefont {Gurarie}}, \bibinfo
  {author} {\bibfnamefont {C.}~\bibnamefont {Xu}}, \bibinfo {author}
  {\bibfnamefont {P.~S.}\ \bibnamefont {Julienne}}, \bibinfo {author}
  {\bibfnamefont {J.}~\bibnamefont {Ye}}, \bibinfo {author} {\bibfnamefont
  {P.}~\bibnamefont {Zoller}}, \bibinfo {author} {\bibfnamefont
  {E.}~\bibnamefont {Demler}}, \bibinfo {author} {\bibfnamefont {M.~D.}\
  \bibnamefont {Lukin}}, \ and\ \bibinfo {author} {\bibfnamefont {A.~M.}\
  \bibnamefont {Rey}},\ }\href {http://dx.doi.org/10.1038/nphys1535} {\bibfield
   {journal} {\bibinfo  {journal} {Nature Physics}\ }\textbf {\bibinfo {volume}
  {6}},\ \bibinfo {pages} {289} (\bibinfo {year} {2010})}\BibitemShut {NoStop}%
\bibitem [{\citenamefont {T\'oth}\ \emph {et~al.}(2010)\citenamefont {T\'oth},
  \citenamefont {L\"auchli}, \citenamefont {Mila},\ and\ \citenamefont
  {Penc}}]{Toth2010}%
  \BibitemOpen
  \bibfield  {author} {\bibinfo {author} {\bibfnamefont {T.~A.}\ \bibnamefont
  {T\'oth}}, \bibinfo {author} {\bibfnamefont {A.~M.}\ \bibnamefont
  {L\"auchli}}, \bibinfo {author} {\bibfnamefont {F.}~\bibnamefont {Mila}}, \
  and\ \bibinfo {author} {\bibfnamefont {K.}~\bibnamefont {Penc}},\ }\href
  {\doibase 10.1103/PhysRevLett.105.265301} {\bibfield  {journal} {\bibinfo
  {journal} {Phys. Rev. Lett.}\ }\textbf {\bibinfo {volume} {105}},\ \bibinfo
  {pages} {265301} (\bibinfo {year} {2010})}\BibitemShut {NoStop}%
\bibitem [{\citenamefont {Sotnikov}\ and\ \citenamefont
  {Hofstetter}(2014)}]{Sotnikov2014}%
  \BibitemOpen
  \bibfield  {author} {\bibinfo {author} {\bibfnamefont {A.}~\bibnamefont
  {Sotnikov}}\ and\ \bibinfo {author} {\bibfnamefont {W.}~\bibnamefont
  {Hofstetter}},\ }\href {\doibase 10.1103/PhysRevA.89.063601} {\bibfield
  {journal} {\bibinfo  {journal} {Phys. Rev. A}\ }\textbf {\bibinfo {volume}
  {89}},\ \bibinfo {pages} {063601} (\bibinfo {year} {2014})}\BibitemShut
  {NoStop}%
\bibitem [{\citenamefont {Hermele}\ \emph {et~al.}(2009)\citenamefont
  {Hermele}, \citenamefont {Gurarie},\ and\ \citenamefont {Rey}}]{Hermele2009}%
  \BibitemOpen
  \bibfield  {author} {\bibinfo {author} {\bibfnamefont {M.}~\bibnamefont
  {Hermele}}, \bibinfo {author} {\bibfnamefont {V.}~\bibnamefont {Gurarie}}, \
  and\ \bibinfo {author} {\bibfnamefont {A.~M.}\ \bibnamefont {Rey}},\ }\href
  {\doibase 10.1103/PhysRevLett.103.135301} {\bibfield  {journal} {\bibinfo
  {journal} {Phys. Rev. Lett.}\ }\textbf {\bibinfo {volume} {103}},\ \bibinfo
  {pages} {135301} (\bibinfo {year} {2009})}\BibitemShut {NoStop}%
\bibitem [{\citenamefont {Hermele}\ and\ \citenamefont
  {Gurarie}(2011)}]{Hermele2011}%
  \BibitemOpen
  \bibfield  {author} {\bibinfo {author} {\bibfnamefont {M.}~\bibnamefont
  {Hermele}}\ and\ \bibinfo {author} {\bibfnamefont {V.}~\bibnamefont
  {Gurarie}},\ }\href {\doibase 10.1103/PhysRevB.84.174441} {\bibfield
  {journal} {\bibinfo  {journal} {Phys. Rev. B}\ }\textbf {\bibinfo {volume}
  {84}},\ \bibinfo {pages} {174441} (\bibinfo {year} {2011})}\BibitemShut
  {NoStop}%
\bibitem [{\citenamefont {Zhou}\ \emph {et~al.}(2016)\citenamefont {Zhou},
  \citenamefont {Wang}, \citenamefont {Meng}, \citenamefont {Wang},\ and\
  \citenamefont {Wu}}]{Zhou2016}%
  \BibitemOpen
  \bibfield  {author} {\bibinfo {author} {\bibfnamefont {Z.}~\bibnamefont
  {Zhou}}, \bibinfo {author} {\bibfnamefont {D.}~\bibnamefont {Wang}}, \bibinfo
  {author} {\bibfnamefont {Z.~Y.}\ \bibnamefont {Meng}}, \bibinfo {author}
  {\bibfnamefont {Y.}~\bibnamefont {Wang}}, \ and\ \bibinfo {author}
  {\bibfnamefont {C.}~\bibnamefont {Wu}},\ }\href {\doibase
  10.1103/PhysRevB.93.245157} {\bibfield  {journal} {\bibinfo  {journal} {Phys.
  Rev. B}\ }\textbf {\bibinfo {volume} {93}},\ \bibinfo {pages} {245157}
  (\bibinfo {year} {2016})}\BibitemShut {NoStop}%
\bibitem [{\citenamefont {Cole}\ \emph {et~al.}(2012)\citenamefont {Cole},
  \citenamefont {Zhang}, \citenamefont {Paramekanti},\ and\ \citenamefont
  {Trivedi}}]{Cole2012}%
  \BibitemOpen
  \bibfield  {author} {\bibinfo {author} {\bibfnamefont {W.~S.}\ \bibnamefont
  {Cole}}, \bibinfo {author} {\bibfnamefont {S.}~\bibnamefont {Zhang}},
  \bibinfo {author} {\bibfnamefont {A.}~\bibnamefont {Paramekanti}}, \ and\
  \bibinfo {author} {\bibfnamefont {N.}~\bibnamefont {Trivedi}},\ }\href
  {\doibase 10.1103/PhysRevLett.109.085302} {\bibfield  {journal} {\bibinfo
  {journal} {Phys. Rev. Lett.}\ }\textbf {\bibinfo {volume} {109}},\ \bibinfo
  {pages} {085302} (\bibinfo {year} {2012})}\BibitemShut {NoStop}%
\bibitem [{\citenamefont {Radi\ifmmode~\acute{c}\else \'{c}\fi{}}\ \emph
  {et~al.}(2012)\citenamefont {Radi\ifmmode~\acute{c}\else \'{c}\fi{}},
  \citenamefont {Di~Ciolo}, \citenamefont {Sun},\ and\ \citenamefont
  {Galitski}}]{Radic2012}%
  \BibitemOpen
  \bibfield  {author} {\bibinfo {author} {\bibfnamefont {J.}~\bibnamefont
  {Radi\ifmmode~\acute{c}\else \'{c}\fi{}}}, \bibinfo {author} {\bibfnamefont
  {A.}~\bibnamefont {Di~Ciolo}}, \bibinfo {author} {\bibfnamefont
  {K.}~\bibnamefont {Sun}}, \ and\ \bibinfo {author} {\bibfnamefont
  {V.}~\bibnamefont {Galitski}},\ }\href {\doibase
  10.1103/PhysRevLett.109.085303} {\bibfield  {journal} {\bibinfo  {journal}
  {Phys. Rev. Lett.}\ }\textbf {\bibinfo {volume} {109}},\ \bibinfo {pages}
  {085303} (\bibinfo {year} {2012})}\BibitemShut {NoStop}%
\bibitem [{\citenamefont {Cocks}\ \emph {et~al.}(2012)\citenamefont {Cocks},
  \citenamefont {Orth}, \citenamefont {Rachel}, \citenamefont {Buchhold},
  \citenamefont {Le~Hur},\ and\ \citenamefont {Hofstetter}}]{Cocks2012}%
  \BibitemOpen
  \bibfield  {author} {\bibinfo {author} {\bibfnamefont {D.}~\bibnamefont
  {Cocks}}, \bibinfo {author} {\bibfnamefont {P.~P.}\ \bibnamefont {Orth}},
  \bibinfo {author} {\bibfnamefont {S.}~\bibnamefont {Rachel}}, \bibinfo
  {author} {\bibfnamefont {M.}~\bibnamefont {Buchhold}}, \bibinfo {author}
  {\bibfnamefont {K.}~\bibnamefont {Le~Hur}}, \ and\ \bibinfo {author}
  {\bibfnamefont {W.}~\bibnamefont {Hofstetter}},\ }\href {\doibase
  10.1103/PhysRevLett.109.205303} {\bibfield  {journal} {\bibinfo  {journal}
  {Phys. Rev. Lett.}\ }\textbf {\bibinfo {volume} {109}},\ \bibinfo {pages}
  {205303} (\bibinfo {year} {2012})}\BibitemShut {NoStop}%
\bibitem [{\citenamefont {Hickey}\ \emph {et~al.}(2015)\citenamefont {Hickey},
  \citenamefont {Rath},\ and\ \citenamefont {Paramekanti}}]{Hickey2015}%
  \BibitemOpen
  \bibfield  {author} {\bibinfo {author} {\bibfnamefont {C.}~\bibnamefont
  {Hickey}}, \bibinfo {author} {\bibfnamefont {P.}~\bibnamefont {Rath}}, \ and\
  \bibinfo {author} {\bibfnamefont {A.}~\bibnamefont {Paramekanti}},\ }\href
  {\doibase 10.1103/PhysRevB.91.134414} {\bibfield  {journal} {\bibinfo
  {journal} {Phys. Rev. B}\ }\textbf {\bibinfo {volume} {91}},\ \bibinfo
  {pages} {134414} (\bibinfo {year} {2015})}\BibitemShut {NoStop}%
\bibitem [{\citenamefont {Barnett}\ \emph {et~al.}(2012)\citenamefont
  {Barnett}, \citenamefont {Boyd},\ and\ \citenamefont
  {Galitski}}]{Barnett2012}%
  \BibitemOpen
  \bibfield  {author} {\bibinfo {author} {\bibfnamefont {R.}~\bibnamefont
  {Barnett}}, \bibinfo {author} {\bibfnamefont {G.~R.}\ \bibnamefont {Boyd}}, \
  and\ \bibinfo {author} {\bibfnamefont {V.}~\bibnamefont {Galitski}},\ }\href
  {\doibase 10.1103/PhysRevLett.109.235308} {\bibfield  {journal} {\bibinfo
  {journal} {Phys. Rev. Lett.}\ }\textbf {\bibinfo {volume} {109}},\ \bibinfo
  {pages} {235308} (\bibinfo {year} {2012})}\BibitemShut {NoStop}%
\bibitem [{\citenamefont {Juzeli\ifmmode~\bar{u}\else \={u}\fi{}nas}\ \emph
  {et~al.}(2010)\citenamefont {Juzeli\ifmmode~\bar{u}\else \={u}\fi{}nas},
  \citenamefont {Ruseckas},\ and\ \citenamefont {Dalibard}}]{Juzeliunas2010}%
  \BibitemOpen
  \bibfield  {author} {\bibinfo {author} {\bibfnamefont {G.}~\bibnamefont
  {Juzeli\ifmmode~\bar{u}\else \={u}\fi{}nas}}, \bibinfo {author}
  {\bibfnamefont {J.}~\bibnamefont {Ruseckas}}, \ and\ \bibinfo {author}
  {\bibfnamefont {J.}~\bibnamefont {Dalibard}},\ }\href {\doibase
  10.1103/PhysRevA.81.053403} {\bibfield  {journal} {\bibinfo  {journal} {Phys.
  Rev. A}\ }\textbf {\bibinfo {volume} {81}},\ \bibinfo {pages} {053403}
  (\bibinfo {year} {2010})}\BibitemShut {NoStop}%
\bibitem [{\citenamefont {Dalibard}\ \emph {et~al.}(2011)\citenamefont
  {Dalibard}, \citenamefont {Gerbier}, \citenamefont
  {Juzeli\ifmmode~\bar{u}\else \={u}\fi{}nas},\ and\ \citenamefont
  {\"Ohberg}}]{Dalibard2011}%
  \BibitemOpen
  \bibfield  {author} {\bibinfo {author} {\bibfnamefont {J.}~\bibnamefont
  {Dalibard}}, \bibinfo {author} {\bibfnamefont {F.}~\bibnamefont {Gerbier}},
  \bibinfo {author} {\bibfnamefont {G.}~\bibnamefont
  {Juzeli\ifmmode~\bar{u}\else \={u}\fi{}nas}}, \ and\ \bibinfo {author}
  {\bibfnamefont {P.}~\bibnamefont {\"Ohberg}},\ }\href {\doibase
  10.1103/RevModPhys.83.1523} {\bibfield  {journal} {\bibinfo  {journal} {Rev.
  Mod. Phys.}\ }\textbf {\bibinfo {volume} {83}},\ \bibinfo {pages} {1523}
  (\bibinfo {year} {2011})}\BibitemShut {NoStop}%
\bibitem [{\citenamefont {Goldman}\ \emph {et~al.}(2014)\citenamefont
  {Goldman}, \citenamefont {Juzeliūnas}, \citenamefont {Öhberg},\ and\
  \citenamefont {Spielman}}]{Goldman2014}%
  \BibitemOpen
  \bibfield  {author} {\bibinfo {author} {\bibfnamefont {N.}~\bibnamefont
  {Goldman}}, \bibinfo {author} {\bibfnamefont {G.}~\bibnamefont
  {Juzeliūnas}}, \bibinfo {author} {\bibfnamefont {P.}~\bibnamefont
  {Öhberg}}, \ and\ \bibinfo {author} {\bibfnamefont {I.~B.}\ \bibnamefont
  {Spielman}},\ }\href {http://stacks.iop.org/0034-4885/77/i=12/a=126401}
  {\bibfield  {journal} {\bibinfo  {journal} {Reports on Progress in Physics}\
  }\textbf {\bibinfo {volume} {77}},\ \bibinfo {pages} {126401} (\bibinfo
  {year} {2014})}\BibitemShut {NoStop}%
\bibitem [{\citenamefont {Song}\ \emph {et~al.}(2008)\citenamefont {Song},
  \citenamefont {Wortis},\ and\ \citenamefont {Atkinson}}]{Song2008}%
  \BibitemOpen
  \bibfield  {author} {\bibinfo {author} {\bibfnamefont {Y.}~\bibnamefont
  {Song}}, \bibinfo {author} {\bibfnamefont {R.}~\bibnamefont {Wortis}}, \ and\
  \bibinfo {author} {\bibfnamefont {W.~A.}\ \bibnamefont {Atkinson}},\ }\href
  {\doibase 10.1103/PhysRevB.77.054202} {\bibfield  {journal} {\bibinfo
  {journal} {Phys. Rev. B}\ }\textbf {\bibinfo {volume} {77}},\ \bibinfo
  {pages} {054202} (\bibinfo {year} {2008})}\BibitemShut {NoStop}%
\bibitem [{\citenamefont {Snoek}\ \emph {et~al.}(2008)\citenamefont {Snoek},
  \citenamefont {Titvinidze}, \citenamefont {Tőke}, \citenamefont {Byczuk},\
  and\ \citenamefont {Hofstetter}}]{Snoek2008}%
  \BibitemOpen
  \bibfield  {author} {\bibinfo {author} {\bibfnamefont {M.}~\bibnamefont
  {Snoek}}, \bibinfo {author} {\bibfnamefont {I.}~\bibnamefont {Titvinidze}},
  \bibinfo {author} {\bibfnamefont {C.}~\bibnamefont {Tőke}}, \bibinfo
  {author} {\bibfnamefont {K.}~\bibnamefont {Byczuk}}, \ and\ \bibinfo {author}
  {\bibfnamefont {W.}~\bibnamefont {Hofstetter}},\ }\href
  {http://stacks.iop.org/1367-2630/10/i=9/a=093008} {\bibfield  {journal}
  {\bibinfo  {journal} {New Journal of Physics}\ }\textbf {\bibinfo {volume}
  {10}},\ \bibinfo {pages} {093008} (\bibinfo {year} {2008})}\BibitemShut
  {NoStop}%
\bibitem [{\citenamefont {{Irsigler}}\ \emph {et~al.}(2018)\citenamefont
  {{Irsigler}}, \citenamefont {{Zheng}},\ and\ \citenamefont
  {{Hofstetter}}}]{Irsigler2018}%
  \BibitemOpen
  \bibfield  {author} {\bibinfo {author} {\bibfnamefont {B.}~\bibnamefont
  {{Irsigler}}}, \bibinfo {author} {\bibfnamefont {J.-H.}\ \bibnamefont
  {{Zheng}}}, \ and\ \bibinfo {author} {\bibfnamefont {W.}~\bibnamefont
  {{Hofstetter}}},\ }\href@noop {} {\bibfield  {journal} {\bibinfo  {journal}
  {ArXiv e-prints}\ } (\bibinfo {year} {2018})},\ \Eprint
  {http://arxiv.org/abs/1806.01598} {arXiv:1806.01598 [cond-mat.quant-gas]}
  \BibitemShut {NoStop}%
\bibitem [{\citenamefont {{Zheng}}\ \emph {et~al.}(2018)\citenamefont
  {{Zheng}}, \citenamefont {{Qin}},\ and\ \citenamefont
  {{Hofstetter}}}]{Zheng2018}%
  \BibitemOpen
  \bibfield  {author} {\bibinfo {author} {\bibfnamefont {J.-H.}\ \bibnamefont
  {{Zheng}}}, \bibinfo {author} {\bibfnamefont {T.}~\bibnamefont {{Qin}}}, \
  and\ \bibinfo {author} {\bibfnamefont {W.}~\bibnamefont {{Hofstetter}}},\
  }\href@noop {} {\bibfield  {journal} {\bibinfo  {journal} {ArXiv e-prints}\ }
  (\bibinfo {year} {2018})},\ \Eprint {http://arxiv.org/abs/1805.10491}
  {arXiv:1805.10491 [cond-mat.dis-nn]} \BibitemShut {NoStop}%
\bibitem [{\citenamefont {Inaba}\ \emph {et~al.}(2010)\citenamefont {Inaba},
  \citenamefont {Miyatake},\ and\ \citenamefont {Suga}}]{Inaba2010}%
  \BibitemOpen
  \bibfield  {author} {\bibinfo {author} {\bibfnamefont {K.}~\bibnamefont
  {Inaba}}, \bibinfo {author} {\bibfnamefont {S.-y.}\ \bibnamefont {Miyatake}},
  \ and\ \bibinfo {author} {\bibfnamefont {S.-i.}\ \bibnamefont {Suga}},\
  }\href {\doibase 10.1103/PhysRevA.82.051602} {\bibfield  {journal} {\bibinfo
  {journal} {Phys. Rev. A}\ }\textbf {\bibinfo {volume} {82}},\ \bibinfo
  {pages} {051602} (\bibinfo {year} {2010})}\BibitemShut {NoStop}%
\bibitem [{\citenamefont {Miyatake}\ \emph {et~al.}(2010)\citenamefont
  {Miyatake}, \citenamefont {Inaba},\ and\ \citenamefont
  {Suga}}]{Miyatake2010}%
  \BibitemOpen
  \bibfield  {author} {\bibinfo {author} {\bibfnamefont {S.-y.}\ \bibnamefont
  {Miyatake}}, \bibinfo {author} {\bibfnamefont {K.}~\bibnamefont {Inaba}}, \
  and\ \bibinfo {author} {\bibfnamefont {S.-i.}\ \bibnamefont {Suga}},\ }\href
  {\doibase 10.1103/PhysRevA.81.021603} {\bibfield  {journal} {\bibinfo
  {journal} {Phys. Rev. A}\ }\textbf {\bibinfo {volume} {81}},\ \bibinfo
  {pages} {021603} (\bibinfo {year} {2010})}\BibitemShut {NoStop}%
\bibitem [{\citenamefont {Inaba}\ and\ \citenamefont {Suga}(2012)}]{Inaba2012}%
  \BibitemOpen
  \bibfield  {author} {\bibinfo {author} {\bibfnamefont {K.}~\bibnamefont
  {Inaba}}\ and\ \bibinfo {author} {\bibfnamefont {S.-i.}\ \bibnamefont
  {Suga}},\ }\href {\doibase 10.1103/PhysRevLett.108.255301} {\bibfield
  {journal} {\bibinfo  {journal} {Phys. Rev. Lett.}\ }\textbf {\bibinfo
  {volume} {108}},\ \bibinfo {pages} {255301} (\bibinfo {year}
  {2012})}\BibitemShut {NoStop}%
\bibitem [{\citenamefont {Bauer}\ \emph {et~al.}(2012)\citenamefont {Bauer},
  \citenamefont {Corboz}, \citenamefont {L\"auchli}, \citenamefont {Messio},
  \citenamefont {Penc}, \citenamefont {Troyer},\ and\ \citenamefont
  {Mila}}]{Bauer2012}%
  \BibitemOpen
  \bibfield  {author} {\bibinfo {author} {\bibfnamefont {B.}~\bibnamefont
  {Bauer}}, \bibinfo {author} {\bibfnamefont {P.}~\bibnamefont {Corboz}},
  \bibinfo {author} {\bibfnamefont {A.~M.}\ \bibnamefont {L\"auchli}}, \bibinfo
  {author} {\bibfnamefont {L.}~\bibnamefont {Messio}}, \bibinfo {author}
  {\bibfnamefont {K.}~\bibnamefont {Penc}}, \bibinfo {author} {\bibfnamefont
  {M.}~\bibnamefont {Troyer}}, \ and\ \bibinfo {author} {\bibfnamefont
  {F.}~\bibnamefont {Mila}},\ }\href {\doibase 10.1103/PhysRevB.85.125116}
  {\bibfield  {journal} {\bibinfo  {journal} {Phys. Rev. B}\ }\textbf {\bibinfo
  {volume} {85}},\ \bibinfo {pages} {125116} (\bibinfo {year}
  {2012})}\BibitemShut {NoStop}%
\bibitem [{\citenamefont {Potthoff}\ and\ \citenamefont
  {Nolting}(1999)}]{Potthoff1999}%
  \BibitemOpen
  \bibfield  {author} {\bibinfo {author} {\bibfnamefont {M.}~\bibnamefont
  {Potthoff}}\ and\ \bibinfo {author} {\bibfnamefont {W.}~\bibnamefont
  {Nolting}},\ }\href {\doibase 10.1103/PhysRevB.59.2549} {\bibfield  {journal}
  {\bibinfo  {journal} {Phys. Rev. B}\ }\textbf {\bibinfo {volume} {59}},\
  \bibinfo {pages} {2549} (\bibinfo {year} {1999})}\BibitemShut {NoStop}%
\bibitem [{\citenamefont {Hofstetter}\ and\ \citenamefont
  {Qin}(2018)}]{Hofstetter2018}%
  \BibitemOpen
  \bibfield  {author} {\bibinfo {author} {\bibfnamefont {W.}~\bibnamefont
  {Hofstetter}}\ and\ \bibinfo {author} {\bibfnamefont {T.}~\bibnamefont
  {Qin}},\ }\href {http://stacks.iop.org/0953-4075/51/i=8/a=082001} {\bibfield
  {journal} {\bibinfo  {journal} {Journal of Physics B: Atomic, Molecular and
  Optical Physics}\ }\textbf {\bibinfo {volume} {51}},\ \bibinfo {pages}
  {082001} (\bibinfo {year} {2018})}\BibitemShut {NoStop}%
\bibitem [{\citenamefont {Orth}\ \emph {et~al.}(2013)\citenamefont {Orth},
  \citenamefont {Cocks}, \citenamefont {Rachel}, \citenamefont {Buchhold},
  \citenamefont {Hur},\ and\ \citenamefont {Hofstetter}}]{Orth2013}%
  \BibitemOpen
  \bibfield  {author} {\bibinfo {author} {\bibfnamefont {P.~P.}\ \bibnamefont
  {Orth}}, \bibinfo {author} {\bibfnamefont {D.}~\bibnamefont {Cocks}},
  \bibinfo {author} {\bibfnamefont {S.}~\bibnamefont {Rachel}}, \bibinfo
  {author} {\bibfnamefont {M.}~\bibnamefont {Buchhold}}, \bibinfo {author}
  {\bibfnamefont {K.~L.}\ \bibnamefont {Hur}}, \ and\ \bibinfo {author}
  {\bibfnamefont {W.}~\bibnamefont {Hofstetter}},\ }\href
  {http://stacks.iop.org/0953-4075/46/i=13/a=134004} {\bibfield  {journal}
  {\bibinfo  {journal} {Journal of Physics B: Atomic, Molecular and Optical
  Physics}\ }\textbf {\bibinfo {volume} {46}},\ \bibinfo {pages} {134004}
  (\bibinfo {year} {2013})}\BibitemShut {NoStop}%
\bibitem [{\citenamefont {He}\ \emph {et~al.}(2015)\citenamefont {He},
  \citenamefont {Ji},\ and\ \citenamefont {Hofstetter}}]{He2015}%
  \BibitemOpen
  \bibfield  {author} {\bibinfo {author} {\bibfnamefont {L.}~\bibnamefont
  {He}}, \bibinfo {author} {\bibfnamefont {A.}~\bibnamefont {Ji}}, \ and\
  \bibinfo {author} {\bibfnamefont {W.}~\bibnamefont {Hofstetter}},\ }\href
  {\doibase 10.1103/PhysRevA.92.023630} {\bibfield  {journal} {\bibinfo
  {journal} {Phys. Rev. A}\ }\textbf {\bibinfo {volume} {92}},\ \bibinfo
  {pages} {023630} (\bibinfo {year} {2015})}\BibitemShut {NoStop}%
\bibitem [{\citenamefont {Hatsugai}\ and\ \citenamefont
  {Kohmoto}(1990)}]{Hatsugai1990}%
  \BibitemOpen
  \bibfield  {author} {\bibinfo {author} {\bibfnamefont {Y.}~\bibnamefont
  {Hatsugai}}\ and\ \bibinfo {author} {\bibfnamefont {M.}~\bibnamefont
  {Kohmoto}},\ }\href {\doibase 10.1103/PhysRevB.42.8282} {\bibfield  {journal}
  {\bibinfo  {journal} {Phys. Rev. B}\ }\textbf {\bibinfo {volume} {42}},\
  \bibinfo {pages} {8282} (\bibinfo {year} {1990})}\BibitemShut {NoStop}%
\bibitem [{\citenamefont {Vanhala}\ \emph {et~al.}(2016)\citenamefont
  {Vanhala}, \citenamefont {Siro}, \citenamefont {Liang}, \citenamefont
  {Troyer}, \citenamefont {Harju},\ and\ \citenamefont
  {T\"orm\"a}}]{Vanhala2016}%
  \BibitemOpen
  \bibfield  {author} {\bibinfo {author} {\bibfnamefont {T.~I.}\ \bibnamefont
  {Vanhala}}, \bibinfo {author} {\bibfnamefont {T.}~\bibnamefont {Siro}},
  \bibinfo {author} {\bibfnamefont {L.}~\bibnamefont {Liang}}, \bibinfo
  {author} {\bibfnamefont {M.}~\bibnamefont {Troyer}}, \bibinfo {author}
  {\bibfnamefont {A.}~\bibnamefont {Harju}}, \ and\ \bibinfo {author}
  {\bibfnamefont {P.}~\bibnamefont {T\"orm\"a}},\ }\href {\doibase
  10.1103/PhysRevLett.116.225305} {\bibfield  {journal} {\bibinfo  {journal}
  {Phys. Rev. Lett.}\ }\textbf {\bibinfo {volume} {116}},\ \bibinfo {pages}
  {225305} (\bibinfo {year} {2016})}\BibitemShut {NoStop}%
\bibitem [{\citenamefont {Zheng}\ \emph {et~al.}(2015)\citenamefont {Zheng},
  \citenamefont {Shen}, \citenamefont {Wang},\ and\ \citenamefont
  {Zhai}}]{Zheng2015}%
  \BibitemOpen
  \bibfield  {author} {\bibinfo {author} {\bibfnamefont {W.}~\bibnamefont
  {Zheng}}, \bibinfo {author} {\bibfnamefont {H.}~\bibnamefont {Shen}},
  \bibinfo {author} {\bibfnamefont {Z.}~\bibnamefont {Wang}}, \ and\ \bibinfo
  {author} {\bibfnamefont {H.}~\bibnamefont {Zhai}},\ }\href {\doibase
  10.1103/PhysRevB.91.161107} {\bibfield  {journal} {\bibinfo  {journal} {Phys.
  Rev. B}\ }\textbf {\bibinfo {volume} {91}},\ \bibinfo {pages} {161107}
  (\bibinfo {year} {2015})}\BibitemShut {NoStop}%
\bibitem [{\citenamefont {Ottenstein}\ \emph {et~al.}(2008)\citenamefont
  {Ottenstein}, \citenamefont {Lompe}, \citenamefont {Kohnen}, \citenamefont
  {Wenz},\ and\ \citenamefont {Jochim}}]{Ottenstein2008}%
  \BibitemOpen
  \bibfield  {author} {\bibinfo {author} {\bibfnamefont {T.~B.}\ \bibnamefont
  {Ottenstein}}, \bibinfo {author} {\bibfnamefont {T.}~\bibnamefont {Lompe}},
  \bibinfo {author} {\bibfnamefont {M.}~\bibnamefont {Kohnen}}, \bibinfo
  {author} {\bibfnamefont {A.~N.}\ \bibnamefont {Wenz}}, \ and\ \bibinfo
  {author} {\bibfnamefont {S.}~\bibnamefont {Jochim}},\ }\href {\doibase
  10.1103/PhysRevLett.101.203202} {\bibfield  {journal} {\bibinfo  {journal}
  {Phys. Rev. Lett.}\ }\textbf {\bibinfo {volume} {101}},\ \bibinfo {pages}
  {203202} (\bibinfo {year} {2008})}\BibitemShut {NoStop}%
\bibitem [{\citenamefont {Huckans}\ \emph {et~al.}(2009)\citenamefont
  {Huckans}, \citenamefont {Williams}, \citenamefont {Hazlett}, \citenamefont
  {Stites},\ and\ \citenamefont {O'Hara}}]{Huckans2009}%
  \BibitemOpen
  \bibfield  {author} {\bibinfo {author} {\bibfnamefont {J.~H.}\ \bibnamefont
  {Huckans}}, \bibinfo {author} {\bibfnamefont {J.~R.}\ \bibnamefont
  {Williams}}, \bibinfo {author} {\bibfnamefont {E.~L.}\ \bibnamefont
  {Hazlett}}, \bibinfo {author} {\bibfnamefont {R.~W.}\ \bibnamefont {Stites}},
  \ and\ \bibinfo {author} {\bibfnamefont {K.~M.}\ \bibnamefont {O'Hara}},\
  }\href {\doibase 10.1103/PhysRevLett.102.165302} {\bibfield  {journal}
  {\bibinfo  {journal} {Phys. Rev. Lett.}\ }\textbf {\bibinfo {volume} {102}},\
  \bibinfo {pages} {165302} (\bibinfo {year} {2009})}\BibitemShut {NoStop}%
\bibitem [{\citenamefont {Sotnikov}(2015)}]{Sotnikov2015}%
  \BibitemOpen
  \bibfield  {author} {\bibinfo {author} {\bibfnamefont {A.}~\bibnamefont
  {Sotnikov}},\ }\href {\doibase 10.1103/PhysRevA.92.023633} {\bibfield
  {journal} {\bibinfo  {journal} {Phys. Rev. A}\ }\textbf {\bibinfo {volume}
  {92}},\ \bibinfo {pages} {023633} (\bibinfo {year} {2015})}\BibitemShut
  {NoStop}%
\bibitem [{\citenamefont {Struck}\ \emph {et~al.}(2011)\citenamefont {Struck},
  \citenamefont {{\"O}lschl{\"a}ger}, \citenamefont {Le~Targat}, \citenamefont
  {Soltan-Panahi}, \citenamefont {Eckardt}, \citenamefont {Lewenstein},
  \citenamefont {Windpassinger},\ and\ \citenamefont {Sengstock}}]{Struck2011}%
  \BibitemOpen
  \bibfield  {author} {\bibinfo {author} {\bibfnamefont {J.}~\bibnamefont
  {Struck}}, \bibinfo {author} {\bibfnamefont {C.}~\bibnamefont
  {{\"O}lschl{\"a}ger}}, \bibinfo {author} {\bibfnamefont {R.}~\bibnamefont
  {Le~Targat}}, \bibinfo {author} {\bibfnamefont {P.}~\bibnamefont
  {Soltan-Panahi}}, \bibinfo {author} {\bibfnamefont {A.}~\bibnamefont
  {Eckardt}}, \bibinfo {author} {\bibfnamefont {M.}~\bibnamefont {Lewenstein}},
  \bibinfo {author} {\bibfnamefont {P.}~\bibnamefont {Windpassinger}}, \ and\
  \bibinfo {author} {\bibfnamefont {K.}~\bibnamefont {Sengstock}},\ }\href
  {\doibase 10.1126/science.1207239} {\bibfield  {journal} {\bibinfo  {journal}
  {Science}\ }\textbf {\bibinfo {volume} {333}},\ \bibinfo {pages} {996}
  (\bibinfo {year} {2011})}\BibitemShut {NoStop}%
\bibitem [{\citenamefont {Duan}\ \emph {et~al.}(2003)\citenamefont {Duan},
  \citenamefont {Demler},\ and\ \citenamefont {Lukin}}]{Duan2003}%
  \BibitemOpen
  \bibfield  {author} {\bibinfo {author} {\bibfnamefont {L.-M.}\ \bibnamefont
  {Duan}}, \bibinfo {author} {\bibfnamefont {E.}~\bibnamefont {Demler}}, \ and\
  \bibinfo {author} {\bibfnamefont {M.~D.}\ \bibnamefont {Lukin}},\ }\href
  {\doibase 10.1103/PhysRevLett.91.090402} {\bibfield  {journal} {\bibinfo
  {journal} {Phys. Rev. Lett.}\ }\textbf {\bibinfo {volume} {91}},\ \bibinfo
  {pages} {090402} (\bibinfo {year} {2003})}\BibitemShut {NoStop}%
\bibitem [{\citenamefont {Hasan}\ and\ \citenamefont {Kane}(2010)}]{Hasan2010}%
  \BibitemOpen
  \bibfield  {author} {\bibinfo {author} {\bibfnamefont {M.~Z.}\ \bibnamefont
  {Hasan}}\ and\ \bibinfo {author} {\bibfnamefont {C.~L.}\ \bibnamefont
  {Kane}},\ }\href {\doibase 10.1103/RevModPhys.82.3045} {\bibfield  {journal}
  {\bibinfo  {journal} {Rev. Mod. Phys.}\ }\textbf {\bibinfo {volume} {82}},\
  \bibinfo {pages} {3045} (\bibinfo {year} {2010})}\BibitemShut {NoStop}%
\bibitem [{\citenamefont {{Del Re}}\ and\ \citenamefont
  {{Capone}}(2017)}]{DelRe2017}%
  \BibitemOpen
  \bibfield  {author} {\bibinfo {author} {\bibfnamefont {L.}~\bibnamefont {{Del
  Re}}}\ and\ \bibinfo {author} {\bibfnamefont {M.}~\bibnamefont {{Capone}}},\
  }\href@noop {} {\bibfield  {journal} {\bibinfo  {journal} {ArXiv e-prints}\ }
  (\bibinfo {year} {2017})},\ \Eprint {http://arxiv.org/abs/1708.00310}
  {arXiv:1708.00310 [cond-mat.quant-gas]} \BibitemShut {NoStop}%
\bibitem [{\citenamefont {Wang}\ and\ \citenamefont {Zhang}(2012)}]{Wang2012}%
  \BibitemOpen
  \bibfield  {author} {\bibinfo {author} {\bibfnamefont {Z.}~\bibnamefont
  {Wang}}\ and\ \bibinfo {author} {\bibfnamefont {S.-C.}\ \bibnamefont
  {Zhang}},\ }\href {\doibase 10.1103/PhysRevX.2.031008} {\bibfield  {journal}
  {\bibinfo  {journal} {Phys. Rev. X}\ }\textbf {\bibinfo {volume} {2}},\
  \bibinfo {pages} {031008} (\bibinfo {year} {2012})}\BibitemShut {NoStop}%
\bibitem [{\citenamefont {Hafez-Torbati}\ and\ \citenamefont
  {Uhrig}(2016)}]{Torbati2016}%
  \BibitemOpen
  \bibfield  {author} {\bibinfo {author} {\bibfnamefont {M.}~\bibnamefont
  {Hafez-Torbati}}\ and\ \bibinfo {author} {\bibfnamefont {G.~S.}\ \bibnamefont
  {Uhrig}},\ }\href {\doibase 10.1103/PhysRevB.93.195128} {\bibfield  {journal}
  {\bibinfo  {journal} {Phys. Rev. B}\ }\textbf {\bibinfo {volume} {93}},\
  \bibinfo {pages} {195128} (\bibinfo {year} {2016})}\BibitemShut {NoStop}%
\end{thebibliography}

%

\end{document}